\documentclass[a4paper]{article}

\usepackage{algorithm, algpseudocode, amsfonts, amsmath, amsopn, amssymb, amsthm, bbm, booktabs, color, enumitem, graphicx, multicol, multirow, rotating, subfig, tikz, url, xcolor, natbib}
\usepackage[a4paper]{geometry}
\usepackage[export]{adjustbox}

\newcommand{\one}{\mathbbm{1}}

\newcommand{\CRPS}{\textrm{CRPS}} 
\newcommand{\BS}{\textrm{BS}}

\newcommand{\MCB}{\textrm{MCB}} 
\newcommand{\DSC}{\textrm{DSC}}
\newcommand{\UNC}{\textrm{UNC}}
\newcommand{\Sbar}{\overline{\rm S}}

\begin{document}

		
\title{Physics-based vs.~data-driven 24-hour probabilistic forecasts of precipitation for northern tropical Africa}

\author{Eva-Maria Walz \thanks{Institute for Stochastics, Karlsruhe Institute of Technology (KIT), Karlsruhe, Germany} \thanks{Computational Statistics (CST) group, Heidelberg Institute for Theoretical Studies, Heidelberg, Germany}  \and Peter Knippertz \thanks{Institute of Meteorology and Climate Research, Karlsruhe Institute of Technology (KIT), Karlsruhe, Germany} \and Andreas H.\ Fink  \footnotemark[3] \and  Gregor Köhler \thanks{German Cancer Research Cancer (DKFZ), Heidelberg, Germany} \and Tilmann Gneiting \footnotemark[2] \footnotemark[1]}
		
		

\maketitle
		
	
\begin{abstract}
Numerical weather prediction (NWP) models struggle to skillfully predict tropical precipitation occurrence and amount, calling for alternative approaches.  For instance, it has been shown that fairly simple, purely data-driven logistic regression models for 24-hour precipitation occurrence outperform both climatological and NWP forecasts for the West African summer monsoon.  More complex neural network based approaches, however, remain underdeveloped due to the non-Gaussian character of precipitation.  In this study, we develop, apply and evaluate a novel two-stage approach, where we first train a U-Net convolutional neural network (CNN) model on gridded rainfall data to obtain a deterministic forecast and then apply the recently developed, nonparametric Easy Uncertainty Quantification (EasyUQ) approach to convert it into a probabilistic forecast.
We evaluate CNN+EasyUQ for one-day ahead 24-hour accumulated precipitation forecasts over northern tropical Africa for 2011--2019, with the Integrated Multi-satellitE Retrievals for GPM (IMERG) data serving as ground truth.  In the most comprehensive assessment to date we compare CNN+EasyUQ to state-of-the-art physics-based and data-driven approaches such as a monthly probabilistic climatology, raw and postprocessed ensemble forecasts from the European Centre for Medium-Range Weather Forecasts (ECMWF), and traditional statistical approaches that use up to 25 predictor variables from IMERG and the ERA5 reanalysis.  Generally, statistical approaches perform about en par with post-processed ECMWF ensemble forecasts.  The CNN+EasyUQ approach, however, clearly outperforms all competitors for both occurrence and amount.  Hybrid methods that merge CNN+EasyUQ and physics-based forecasts show slight further improvement.  Thus, the CNN+EasyUQ approach can likely improve operational probabilistic forecasts of rainfall in the tropics, and potentially even beyond.
\end{abstract}
	
\section{Introduction}  \label{sec:introduction}

Despite the continuous improvement of numerical weather prediction (NWP) models, precipitation forecasts in the tropics remain a great challenge.  Several studies \citep{Haiden.2012, Vogel.2020} have shown that NWP models have difficulties in outperforming climatological forecasts.  A possible explanation is the exceptional high degree of convective organization over tropical Africa \citep{Nesbitt2006, Roca2014}, a process that is difficult to capture with the convective parameterization of NWP models \citep{Vogel.2018}, although recent developments show some promise \citep{Becker2021}.  Statistical postprocessing, spatial averaging, or temporal aggregation lead to improvements in the skill of raw NWP ensemble grid point forecasts in tropical Africa \citep{Vogel.2020, Stellingwerf2021, Gebremichael2022, Ageet2023}, yet in regions of particularly poor performance of the operational forecast systems, viz.~West and Central Equatorial Africa, the forecast gain over climatology is limited.     

The overall poor performance of current operational systems motivates the development of alternative approaches.  \citet{Vogel.2020} implement a fairly simple purely data-driven logistic regression model for 24-hour precipitation occurrence, which outperforms climatology and NWP forecasts for the summer monsoon season in West Africa.  The predictor variables are designed by exploiting spatial-temporal coherence patterns as developed and investigated further in \citet{Athul2023}.  To this end, the rainfall at each grid point is correlated with the rainfall at all other locations from 1, 2, and 3 days before using the coefficient of predictive ability (CPA) measure \citep{Gneiting.2022}.  The locations showing highest CPA for 1, 2, and 3 days before, respectively, are selected as predictor variables in the logistic regression model.  The good performance of this simple logistic model, which is related to coherent, tropical wave driven spatial propagation of precipitation features in West Africa \citep{Athul2023}, motivates the development of more sophisticated data-driven models and the usage of additional weather quantities linked to rainfall occurrence and amount. 

\citet{Vogel.2021} and \citet{Athul2023} have only investigated the skill of probability forecasts for the binary problem of precipitation occurrence.  In this paper, the more challenging problem of producing accurate probabilistic forecasts for accumulated precipitation, a non-negative real-valued variable, is considered.   Precipitation accumulation is generally considered the ``most difficult weather variable to forecast'' \citep{EbertUphoff2023}.  Indeed, precipitation accumulation follows a mixture distribution with a point mass at zero --- namely, for no precipitation --- and a continuous part on the positive real numbers.  Therefore, despite the sweeping rise of data-driven weather prediction \citep{Bouallegue2023} and rapid progress in data-driven nowcasting of precipitation \citep{Ayzel2020, Lagerquist2021, Ravuri2021, Witt.2021, Espeholt2022, Zhang2023}, the development of machine learning based methods for probabilistic quantitative precipitation forecasts --- at least for times larger than 12 hours --- has been lagging.  For example, precipitation was ``not investigated'' \citep[p.~537]{Bi2023} by the Pangu-Weather team and ``left out of the scope" of the GraphCast development, because ``precipitation is sparse and non-Gaussian and would have possibly required different modeling decisions than the other variables'' \citep[p.~1421]{Lam2023}. We address these challenges by developing a novel two-stage CNN+EasyUQ approach, where we first train a U-Net convolutional neural network (CNN) model to obtain a single-valued deterministic forecast, and then use the Easy Uncertainty Quantification (EasyUQ) approach developed by \citet{Walz2023} to convert the deterministic forecast into a probabilistic forecast.

The paper is structured as followed.  Section \ref{sec:data_methods} introduces the data used in the analysis.  Then, an overview of weather quantities which are known to be linked to precipitation and thus are candidates for predictor variables is provided in section \ref{sec:predictors}.  Different types of forecasting models are described in section \ref{sec:fct_model}.  Importantly, we compare the CNN+EasyUQ forecasts to a comprehensive suite of state of the art methods that include physics-based raw NWP ensemble forecasts, postprocessed NWP forecasts, data-driven statistical forecasts based on logistic regression and distributional index models (DIMs), and  combined statistical-dynamical (hybrid) approaches. Results from this comparison are presented in section \ref{sec:evaluation} with the main conclusion and outlook in section \ref{sec:conclusion}.  
		
\section{Data and methods}  \label{sec:data_methods}

In this study, we use data from three different sources.  The arguably best currently available high-resolution, gauge-calibrated, gridded precipitation product, the Integrated Multi-Satellite Retrievals for GPM (Global Precipitation Measurement) \citep[IMERG;][]{huffman2020}, serves as ground truth for precipitation.  The European Centre for Medium-Range Weather Forecasts (ECMWF) Reanalysis Version 5 \citep[ERA5;][]{Hersbach.2020} product is used to obtain estimates of other weather quantities.  Finally, NWP forecasts, namely the high resolution (HRES) run and the full ECMWF ensemble prediction system (EPS) are downloaded from ECMWF's Meteorological Archival and Retrieval System (MARS; \url{https://www.ecmwf.int/en/forecasts/access-forecasts/access-archive-datasets}). 

\begin{figure}[t]
\centering
\includegraphics[width = \textwidth]{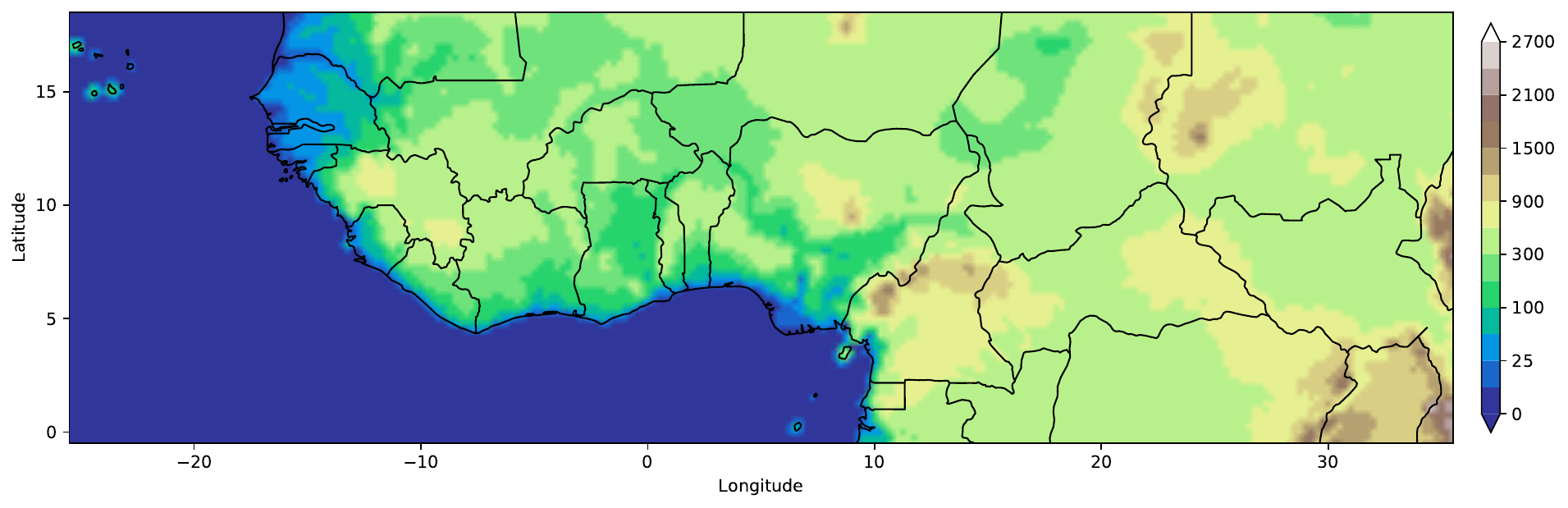}
\caption{Overview of the study area.  Following \citet{Athul2023}, we consider an evaluation domain over northern tropical Africa that comprises $19 \times 61$ grid boxes with centers spanning from $0^\circ$ to $18^\circ$ N in latitude and $25^\circ$ W to $35^\circ$ E in longitude, respectively.  The time period considered ranges from 1 December 2000 to 30 November 2019, with 24-hour forecasts of precipitation amount and precipitation occurrence for 1 December 2010 to 30 November 2019 being evaluated.  The analysis is over land only, and shading indicates altitude in meters, based on the ERA5 land--sea mask. \label{fig:area}}
\end{figure}

The evaluation domain, visualized in Figure~\ref{fig:area}, is northern tropical Africa, represented by $19 \times 61$ grid boxes centered at $0^{\circ}$ to $18^{\circ}$N and $25^{\circ}$W to $35^{\circ}$E, respectively, similar to the setup in \citet{Vogel.2020} and \citet{Athul2023}.  Five distinct seasons are considered as identified previously \citep{fink2017_ch1, Marananetal2018}: December--February (DJF), which is the dry season with occasional showers along the Guinea Coast; the March--April (MA) period, which features highly organized Mesoscale Convective Systems (MCSs) at the Guinea coast and the coastal hinterland; May--June (MJ), the major rainy season along most parts of the Guinea Coast; July--September (JAS), the major rainy season in the Sahel and the little dry season at the coast; and October--November (ON), the second, weaker rainy season at the Guinea Coast.  To avoid cutting seasonal periods at the beginning or the end of the time period under investigation, the time period considered starts 1 December 2000 and ends 30 November 2019.  Importantly, the analysis and evaluation are performed over land only, and we frequently identify a grid box with the grid point at its center.  From now on, when we refer to grid boxes or grid points, we only mean boxes or points on land.

\subsection{GPM IMERG rainfall data}  \label{sec:IMERG}
	
We use the GPM IMERG V06B final version \citep{hou2014, huffman2020} to calculate 24-hour accumulated precipitation from 06 -- 06 UTC for the period under investigation.  GPM IMERG has a temporal resolution of 30 minutes and a spatial resolution of $0.1^\circ \times 0.1^\circ$.  The data were regridded to a resolution of $1^\circ \times 1^\circ$ using first-order conservative remapping.  As we also consider 24-hour rainfall occurrence, we threshold at 0.2 mm to obtain a binary event variable representing precipitation occurrence.

The GPM IMERG algorithm uses both radar-calibrated microwave radiance from polar-orbiting satellites and infrared radiance from geostationary satellites.  In the final version, the precipitation totals are calibrated with rain gauge measurements provided by the Global Precipitation Climatology Centre \citep[GPCC;][]{Schneider2008}.  The degree to which the original estimates are adjusted by the gauge calibration process within a given region is generally determined by the number of available rain gauges, which is highly variable across the Tropics.   
 
\subsection{Predictor variables from ERA5}  \label{sec:ERA5}

Our study considers a range of meteorological variables, specified in section \ref{sec:preds_era5}, as predictor variables for statistical models.  Specifically, we use the ERA5 reanalysis \citep{Hersbach.2020}, which provides a complete and consistent coverage of the study domain by combining model data with observations.  For this study the resolution of the data is $1^\circ \times 1^\circ$ just like for GPM IMERG.  In contrast to 24-hour accumulated precipitation, the considered ERA5 weather quantities are instantaneous values at 00 UTC, thus six hours before the 24-hour accumulation period for GPM IMERG starts.  This way, observed ambient conditions well before the rainfall begins get considered.  For an operational implementation of the respective statistical methods, operational analysis data would need to be used, as ERA5 is not available in near-real time, but we do not expect this to make a big difference to our results.
		
\subsection{Physics-based forecasts from ECMWF}  \label{sec:ECMWF}
     
We now describe the NWP forecasts used in this study, namely, the ECMWF high resolution (HRES) model and ensemble prediction system \citep[EPS;][]{Molteni1996}.  Owing to the high resolution and the initialization with the most accurate analysis product, the HRES model is arguably the leading global deterministic NWP forecast available.  As an operational product, HRES has changed considerably over time in frequent updates (\url{https://confluence.ecmwf.int/display/FCST/Changes+to+the+forecasting+system}).  The ECMWF EPS consists of one control run and 50 perturbed members.  Like the HRES model, the control run is based on the most accurate initial state of the atmosphere.  The perturbed members start from slightly different initial conditions and use perturbed physics options.

The forecasts are available from MARS in a grid resolution of $0.25^\circ \times 0.25^\circ$ and are first-order conservatively remapped to a resolution of $1^\circ \times 1^\circ$.  HRES forecasts for total precipitation are obtained by summing forecasts for large scale precipitation and convective precipitation, which are available from April 2001 on.  For the EPS, total precipitation is available from April 2006 on.  To cover an equal number of seasons, we use data starting in December 2001 and December 2006, respectively.  To obtain forecasts for 24-hour precipitation amount the difference between forecasts of accumulated precipitation initialized at 00 UTC with lead times of 30 and 6 hours is computed.  To compute the EPS forecast probability for the occurrence of precipitation, the member forecasts are thresholded at 0.2 mm and the respective binary outcomes are averaged.

\subsection{Easy Uncertainty Quantification (EasyUQ)}  \label{sec:EasyUQ}

Forecasts ought to take the form of probability distributions to account for uncertainty.  In NWP probabilistic forecasts have become common practice with the operational implementation of ensemble systems \citep{Molteni1996, Bauer2015}.  To quantify uncertainty in very general settings, \citet{Walz2023} introduced EasyUQ, an easy-to-implement method which transforms real-valued deterministic model output into calibrated statistical distributions.  EasyUQ is trained on pairs of deterministic forecasts and corresponding outcomes and is thus independent of the type of model used to generate the single-valued forecasts.  In particular, EasyUQ can be applied to the output of any NWP, statistical, or machine learning model that generates deterministic forecasts.  The EasyUQ forecast distributions are discrete and have mass exclusively at outcome values in the training set.  Therefore, the forecast distributions adapt naturally to the specifics of precipitation accumulation, which follows a mixture distribution with a point mass at zero and a continuous part on the positive real numbers,\footnote{Typically, rainfall amounts are reported in small but fixed increments, so strictly speaking, the distribution on the positive real numbers is discrete as well.  The EasyUQ technique adapts to the level of discretization in the observational record at hand, without any need for user intervention.  However, if continuous forecast distributions on the positive real axis are desirable, adaptations of the Smooth EasyUQ technique developed by \citet{Walz2023} can be employed.} without any need for tuning.

In its basic form, which we use in this study, EasyUQ is a special case of Isotonic Distributional Regression \citep[IDR;][]{Henzi.2021}.  In contrast to NWP ensemble systems, which have large computational costs and require the use of supercomputers \citep{Bauer2015}, the application of EasyUQ to deterministic model output has obvious advantages in terms of the efficient usage of computational resources \citep{Walz2023}. 

\subsection{Evaluation metrics}  \label{sec:metrics}

In our study, we compare methods for probabilistic forecasts of precipitation amount, where the outcome is real-valued, and probability forecasts of precipitation occurrence, where the outcome is binary.  In both cases, we follow extant practice and use proper scoring rules \citep{Gneiting_Raftery.2007}. 

In the setting of probability forecasts, we use the Brier score (BS) to quantify predictive performance based on a collection of pairs $(p_1, y_1), \dots, (p_n, y_n)$ of predictive probabilities and associated binary outcomes.  Specifically, we compute the mean score 
\begin{align}  \label{eq:BS}
\overline{\BS} = \frac{1}{n} \sum_{i=1}^n \BS(p_i,y_i) = \frac{1}{n} \sum_{i=1}^n (p_i-y_i)^2. 
\end{align}
In the case of precipitation amount, we use the continuous ranked probability score (CRPS) for an assessment based on a collection of pairs $(F_1, y_1), \dots, (F_n, y_n)$ of probabilistic forecasts and associated real-valued outcomes.  Comparisons are in terms of the mean score
\begin{align}  \label{eq:CRPS} 
\overline{\CRPS} = \frac{1}{n} \sum_{i=1}^n \CRPS(F_i,y_i) = \frac{1}{n} \sum_{i=1}^n \int_{-\infty}^\infty (F_i(z) - \one \{ z \geq y_i \})^2 \, \textrm{d}z,
\end{align}
where $F_i$ is interpreted as a cumulative distribution function.  To facilitate the assessment of forecast performance relative to a baseline, skill scores can be used, defined as the quantity $(\Sbar_{\text{base}} - \Sbar_{\text{fcst}}) / \, \Sbar_{\text{base}}$, where $\Sbar_{\text{fcst}}$ is the mean score of the forecast at hand and $\Sbar_{\text{base}}$ is the mean score of the baseline.  A positive (negative) Brier or CRPS skill score corresponds to predictive performance better (worse) than the baseline.

For a more informative, diagnostic comparison between forecast methods, we apply the CORP decomposition of \citet{Dimitriadis2021} and the isotonicity-based decomposition of \citet{Arnold2023} to a mean score $\overline{\textrm S}$ from \eqref{eq:BS} or \eqref{eq:CRPS}, respectively.  The decompositions express $\overline{\textrm S}$ in terms of interpretable components, in that
\begin{align}  \label{eq:decomposition}
\overline{\textrm S} = \MCB - \DSC + \UNC, 
\end{align}
where the miscalibration (MCB) component quantifies the (lack of) calibration or reliability of the forecasts (the lower, the better), and the discrimination (DSC) term refers to the discrimination ability or resolution of the forecasts (the higher, the better), whereas the uncertainty (UNC) component is independent of the forecasts and a property of the outcomes only.  For details, we refer to the original work of \citet{Dimitriadis2021} and \citet{Arnold2023}.

\section{Predictor variables for statistical forecasts}  \label{sec:predictors}

In this section we discuss and analyze potential predictor variables for data-driven statistical forecasting methods.  We distinguish predictor variables computed from IMERG data based on spatio-temporal rainfall correlation, and predictor variables based on ERA5.  The initial selection of the variables stems from meteorological expertise.

\subsection{Correlated rainfall predictors from IMERG}  \label{sec:corrs}

\citet{Vogel.2021} introduced a logistic regression model to produce probability forecasts for the binary outcome of precipitation occurrence.  As predictors, they used precipitation data with a lag of one and two days at locations with maximum positive and minimum negative Spearman's rank correlation coefficient.  \citet{Athul2023} noted that due to propagating rainfall systems positive dependencies carry the most useful information, occasionally reaching three days backwards in time.  Moreover, they suggested a replacement of Spearman's rank correlation coefficient by the recently developed coefficient of predictive ability \citep[CPA;][]{Gneiting.2022} measure.  In general, CPA is asymmetric, with the predictor variable and the outcome taking clearly identified roles, as for the classical Area Under the Receiver Operating Characteristic (ROC) Curve (AUC) measure, to which CPA reduces when the outcomes are binary.  When both the predictor variable and the outcome are continuous variables, CPA becomes symmetric and equals Spearman's rank correlation coefficient, up to a linear transformation \citep{Gneiting.2022}.  AUC or CPA values above 0.5 correspond to positive dependencies, and values below 0.5 to negative dependencies. 

Given these insights, this current study uses three correlated precipitation predictor variables, by identifying grid points with maximum CPA at temporal lags of one, two, and three days.  Following \citet{Athul2023}, correlated locations are identified within an enlarged region that comprises $68^{\circ}$W to $50^{\circ}$E and $0^{\circ}$ to $20^{\circ}$N, as compared to the evaluation domain depicted in Figure~\ref{fig:area}, which ranges from $25^{\circ}$W to $35^{\circ}$E and $0^{\circ}$ to $18^{\circ}$N.

\subsection{Predictor variables from ERA5 reanalysis}  \label{sec:preds_era5}

In addition to the correlated precipitation information, various meteorological variables from ERA5 are considered as predictors (Table~\ref{tab:era5_variables}).  For a summary of how environmental conditions affect convection, see \citet{Marananetal2018}.  Unless noted otherwise, the variables are instantaneous quantities at 00 UTC.  The first four variables in Table~\ref{tab:era5_variables} are vertically integrated measures of water in different forms.  TCWV has been shown to be a promising predictor for precipitation by \citet{lafore2017_ch3}; \citet{Witt.2021} use cloud information such as TCLW and TCC in their global statistical model.  The second group comprises the three classical measures of convective instability; CAPE (the theoretical maximum of thermodynamic energy that can be converted into kinetic energy of vertical motion), CIN (the energy barrier that needs to be overcome to reach the level of free convection), and KX (based on dry static vertical stability in the 850--500 hPa layer, absolute humidity at 850 hPa, and relative humidity at 700h pa).  CAPE and CIN have a complex relationship with precipitation and should be considered together and in concert with other parameters \citep{lafore2017_ch11}.  \citet{Galvin2010} demonstrates the usefulness of KX in assessing convective rainfall probability in relation to African Easterly Waves (AEWs). 

\renewcommand{\arraystretch}{1.1}
\begin{table}[t]
\small
\centering
\caption{Predictor variables from ERA5, all at 00 UTC.  \label{tab:era5_variables}}
\begin{tabular}{ll}
\toprule
Meteorological Variable                   & Acronym \\   
\midrule
Total column water vapour                 & TCWV \\      
Vertically integrated moisture divergence & VIMD \\      
Total column cloud liquid water           & TCLW \\      
Total cloud cover                         & TCC \\       
\midrule
Convective available potential energy     & CAPE \\      
Convective inhibition                     & CIN \\       
K-index                                   & KX \\        
\midrule
2m temperature                            & 2T \\        
2m dewpoint temperature                   & 2D \\        
24h surface pressure tendency             & SPT \\       
\midrule
Temperature at 850 hPa                    & T850 \\      
Temperature at 500 hPa                    & T500 \\      
Specific humidity at 925 hPa              & Q925 \\      
Specific humidity at 700 hPa              & Q700 \\      
Specific humidity at 600 hPa              & Q600 \\      
Specific humidity at 500 hPa              & Q500 \\      
Relative humidity at 500 hPa              & R500 \\      
Relative humidity at 300 hPa              & R300 \\      
\midrule
Shear                                     & SHR \\       
Streamfunction at 700 hPa                 & $\Psi$700 \\ 
\bottomrule
\end{tabular}
\end{table}

The third group (2T, 2D, SPT) represents near-surface conditions.  The former two are closely related to the equivalent potential temperature of a starting convective air parcel, thereby influencing the level of cumulus condensation and free convection and thus CIN and CAPE, and have been shown to impact the intensity of convection in West Africa \citep{NichollsMohr2010}.  SPT, the tendency from 00 UTC of the day for which the prediction is made to 00 UTC of the previous day, can be related to AEW propagation and rainfall \citep{Regula1936, Hubert1939}.  The fourth group characterises thermodynamic conditions in the boundary layer and free troposphere between 925 hPa and 300 hPa.  For temperature, we consider 850 hPa and 500 hPa representing lower-tropospheric stability (as in KX).  As moisture generally shows complex vertical structures, 925, 700, 600, and 500 hPa are chosen for specific humidity.  For relative humidity, the mid- to upper-tropospheric levels of 500 hPa and 300 hPa were selected to indicate deep moistening, which facilitates cloud formation and reduces detrimental effects of entrainment on convective development.  Mid-tropospheric relative humidity controls both rainfall enhancement by slow moving tropical waves \citep{Schlueter.2019a} and evaporation of rainfall, and thus convective downdrafts and mesoscale organization of convection \citep{Klein2021}.  The last two entries in Table~\ref{tab:era5_variables} are the circulation-related variables SHR (normalized difference of horizontal wind at 600 and 925 hPa) and $\Psi$700 representing mid-tropospheric streamlines.  SHR influences the potential for mesoscale organization and longevity through separating the areas of convective up- and downdrafts as well as the generation of cold pools \citep{Rotunnoetal1988, lafore2017_ch3}.  Anomalies in $\Psi$700 indicate variations in the African Easterly Jet (AEJ), e.g., passages of troughs and ridges of AEWs \citep{kiladisetal2006}.

\subsection{Statistical analysis of predictor variables}  \label{sec:analysis}

\begin{figure}[t]
\includegraphics[width = \textwidth]{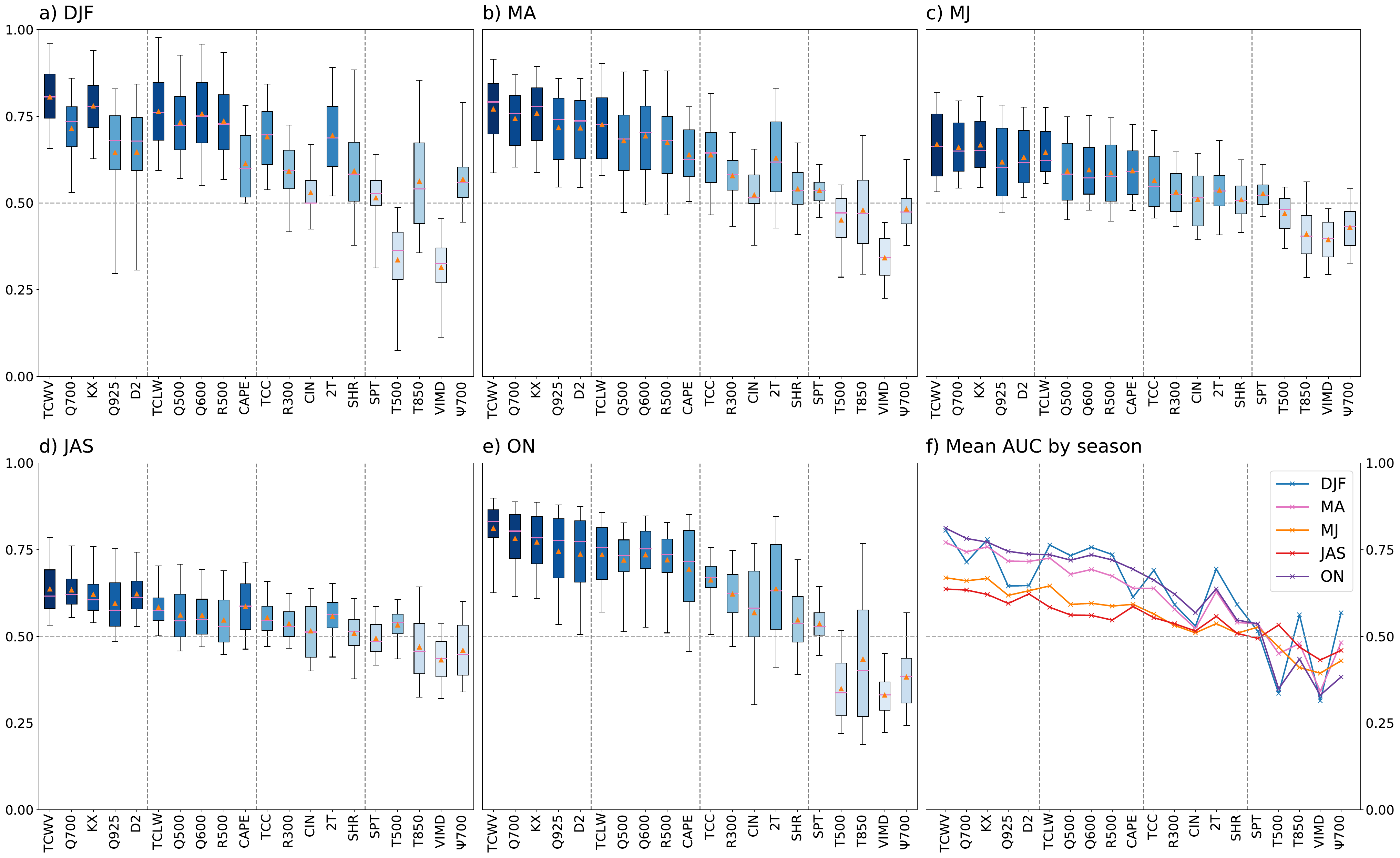}
\caption{Boxplots of grid point AUC values between ERA5 variables from Table \ref{tab:era5_variables} and precipitation occurrence in season a) DJF, b) MA, c) MJ, d) JAS, and e) ON.  The arrangement of the predictor variables on the horizontal axis is in the order of the spatially averaged CPA value for precipitation accumulation, when CPA is computed without splitting into seasons.  The orange marks and the line plots in panel f) indicate the mean AUC value over grid points for the season at hand.  The box colour from dark to light blue indicates the ranking of the seasonal mean AUC value.  In combination this allows to identify differences between yearly vs.\ seasonal perspectives.  \label{fig:AUC}}
\end{figure}

\begin{figure}[t]
\includegraphics[width = \textwidth]{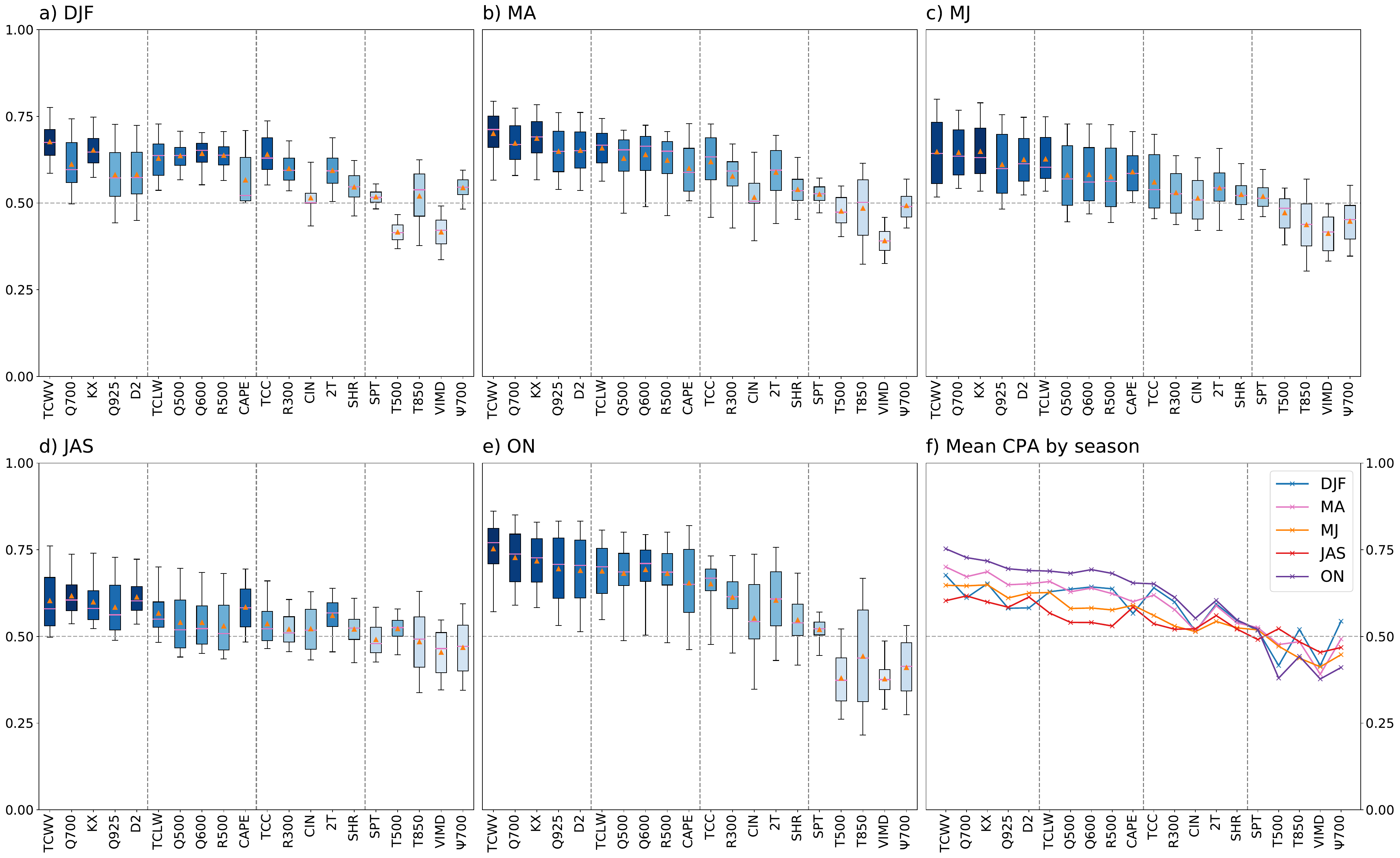}
\caption{As Figure~\ref{fig:AUC} but for CPA and precipitation amount.  \label{fig:CPA}}
\end{figure}

Thus far, the selection of predictor variables has been based on meteorological expertise and findings from other publications.  Here, we use the aforementioned AUC (for rainfall occurrence) and CPA (for amount) measures of \citet{Gneiting.2022} (see Section~\ref{sec:corrs}) for a deeper analysis.  In Figures~\ref{fig:AUC} and \ref{fig:CPA} we show AUC and CPA values for the 20 ERA5 variables from Table~\ref{tab:era5_variables}.  Both are computed in a co-located fashion for each grid point in the evaluation domain (Figure~\ref{fig:area}) and the resulting distributions are represented by boxplots. 

Figure \ref{fig:AUC}a shows AUC values for the dry season DJF.  Given the overall low precipitation amount during this period, the box plots often stretch over large ranges, indicating marked differences between grid points, and also large differences between the variables.  Stable positive relations (i.e., AUC above 0.5) are found for moisture (TCWV, Q500, Q600, Q700, R500), cloud (TCLW, TCC), and instability variables (KX, CAPE), demonstrating a clear dependence on mid-tropospheric conditions, while low-level (Q925, 2D) and upper-level (R300) variables show a more ambiguous behavior.   Other well-defined relations are positive with 2T, and negative with T500 and VIMD.  As the variables are taken at 00 UTC, the relation to 2T may reflect warmer nights under moister and cloudier skies.  CIN, SPT, and $\Psi$700 show weak AUC values close to 0.5.  AUC values for T850 cover a wide range and stretch across 0.5, indicating that its impact depends strongly on the situation.

\begin{figure}[t]
\includegraphics[width = \textwidth]{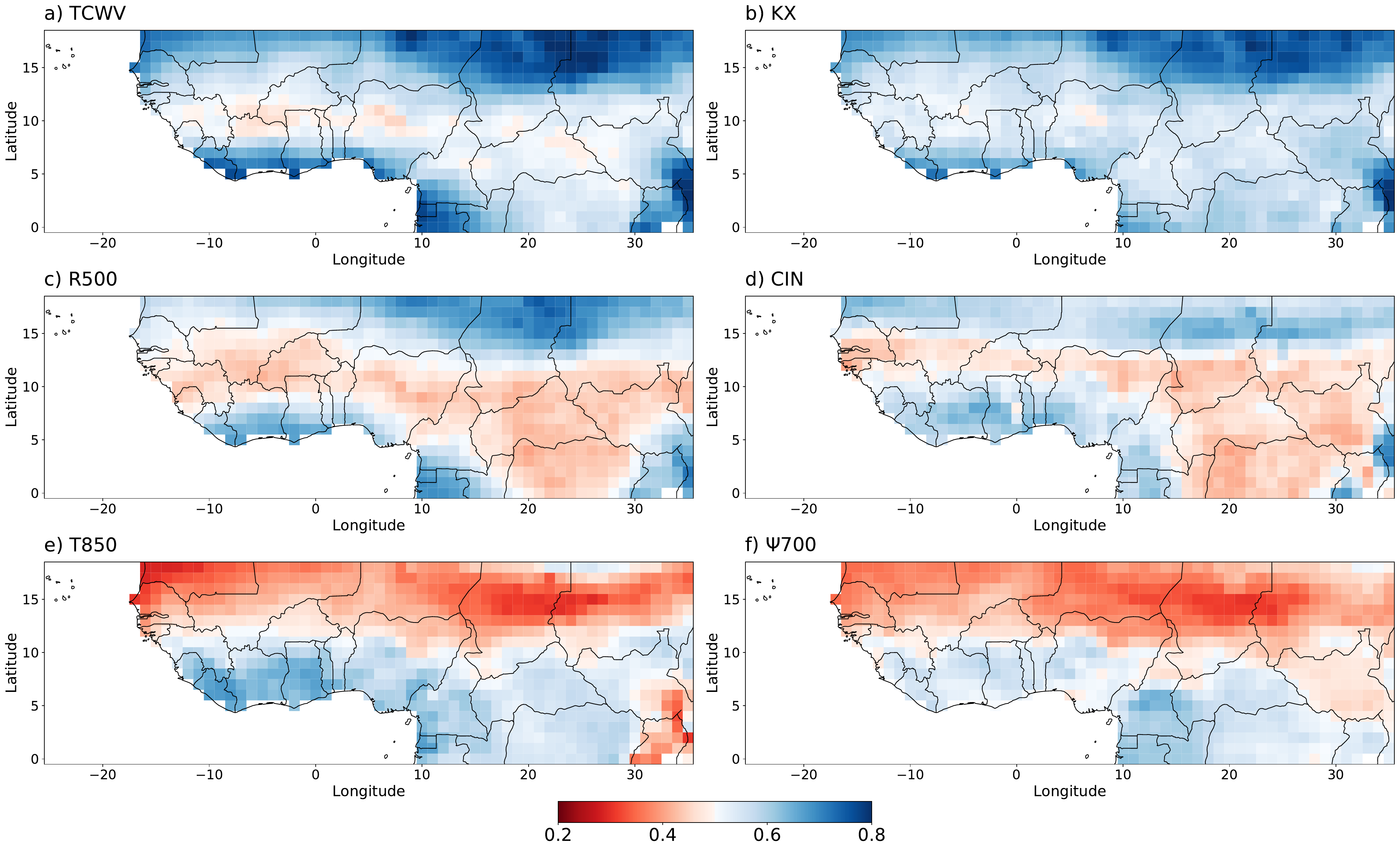}
\caption{Spatial pattern of CPA between the ERA5 predictor a) TCWV, b) KX, c) R500, d) CIN, e) T850, and f) $\Psi$700 from Table \ref{tab:era5_variables} and precipitation amount in season JAS.  \label{fig:CPA_JAS}}
\end{figure}

The corresponding analysis for MA (Figure~\ref{fig:AUC}b) shows an overall less noisy behavior and AUC values more in line with the spatially averaged annual value of CPA that determines the order of the variables in all panels of Figures~\ref{fig:AUC} and \ref{fig:CPA}.  Compared to DJF, a more stable relation to low-level moisture (Q700, Q925, 2D) is visible.  There is a stronger relation to CAPE with little changes in CIN.  Other remarkable changes are less dependence on cold T500, and even more ambiguous relations to T850 and $\Psi$700.  The pre-monsoon season MJ (Figure~\ref{fig:AUC}c), when rainfalls begin to move inland, shows many similarities to MA but the point-to-point variability is smaller and AUC values tend to be closer to 0.5, while their order mostly agrees to that based on annual CPAs.  Remarkable differences to MA are less dependence on 2T and clearer relations to T850 and $\Psi$700 ($< 0.5$).  The latter may indicate a dependence of rainfall on the existence of cyclonic perturbations such as AEWs.  The general magnitude of AUC values close to 0.5 is likely a reflection of the overall improved conditions for convection, which makes individual storms less dependent on particular circumstances, thereby creating a higher degree of stochasticity (see also discussion in \citet{Athul2023}).  This trend continues going into the main monsoon season JAS (Figure~\ref{fig:AUC}d), when most variables show AUC values close to 0.5. The narrower boxplots indicate less local variability during a period when rains penetrate deeply into the continent.  As expected, in the post-monsoon season ON (Figure~\ref{fig:AUC}e), conditions resemble those discussed for MA (Figure~\ref{fig:AUC}b), even with slightly larger amplitudes.  Remarkable differences to MA are that rainfall occurrence depends more on CIN and 2T, possibly because in ON the solar angle is already flatter and the daytime heating is further dampened by the higher moisture availability after the rainy season.  As for DJF, rain depends on cold T500 and the relation to T850 is highly variable and can take both directions, however, with a clear tendency to cooler conditions when rain occurs.  ON also shows the clearest relation to cyclonic perturbations as reflected in AUC values below 0.5 for $\Psi$700.  These may grow in importance relative to other mechanisms, as triggering by daytime heating weakens.  Finally, Figure~\ref{fig:AUC}f shows a summary plot of mean AUC values for all five seasons.  This plot underlines the similar behavior of MJ and JAS (with a consistently higher amplitude for MJ), as well as of MA and ON (with a consistently higher amplitude for ON).  DJF often shows the highest magnitude, as rain depends strongly on unusual conditions to occur, but given the many dry days, the overall behavior appears quite noisy.

The corresponding analysis for CPA is shown in Figure~\ref{fig:CPA}.  Overall there are many similarities to Figure~\ref{fig:AUC}, indicating that variables that work as predictors for occurrence also work for amount.  This is particularly true for the wet part of the year (MJ, JAS and ON), where plots look largely identical (Figure~\ref{fig:CPA}c--e).  For MA (Figure~\ref{fig:CPA}b), there is still large agreement across all variables but the magnitude of CPA values is smaller and the box plots are narrower than for AUC.  This indicates that in this somewhat marginal rainfall season, amount is harder to predict than occurrence.  This trend is even more evident for the dry DJF season (Figure~\ref{fig:CPA}a), when some boxplots become very narrow and magnitudes fall underneath those of ON on average, as shown by the summary plot (Figure~\ref{fig:CPA}f).

For an improved understanding of the ranges in the boxplots, Figure~\ref{fig:CPA_JAS} shows the spatial pattern of CPA of selected meteorological variables exemplarily for the peak monsoon season JAS.  Consistent with the leftmost boxplot in Figure~\ref{fig:CPA}d, CPA values for TCWV are at or above 0.5 almost everywhere in the study region (Figure~\ref{fig:CPA_JAS}a), while featuring an interesting three-tier structure.  Over northern parts of the domain, where moisture is a general limiting factor, CPA values are high, especially over the dry eastern Sahel.  Further south, along the main rain belt and stretching into the Congo Basin, CPA values are close to 0.5, indicating limitations through convective triggering or stability rather than moisture availability.  To the south of the rain belt, i.e., along the Guinea Coast and over the East African highlands, moisture appears to become a limiting factor again.  A very similar pattern but with a smaller range emerges for KX (Figure~\ref{fig:CPA_JAS}b).  The largest differences to TCWV are found along the Guinea Coast, where conditions are often close to moist neutral requiring lifting mechanism to produce rain \citep[cf.~Figure~1.31 in][]{fink2017_ch1}.  Similar but slightly northward shifted structures are found for CAPE (Appendix \ref{app:figures}, Figure~\ref{fig:CPA_JAS_2}c).

A much larger range (0.35--0.75) but with a similar three-tier structure is found for R500 and CIN (Figure~\ref{fig:CPA_JAS}c,d).  One would expect that a moister mid-troposphere and less convective inhibition (recall that CIN is negatively oriented) enhances rainfall amounts and so the behavior within the rain belt is somewhat counter-intuitive.  The most likely explanation is that in areas of abundant moisture and often neutral stratification, large rainfall amounts can most effectively be generated by organized convective systems that require some barrier to accumulate CAPE over the following day and a relatively dry mid-troposphere to allow rainfall evaporation and downdrafts, which in turn can trigger new convection through cold pools (cf.\ Table 11.2 in \citet{lafore2017_ch11}).  It is interesting to note that CPA for TCC is similarly structured as R500, showing CPA values well below 0.5 in the rain belt (Appendix \ref{app:figures}, Figure~\ref{fig:CPA_JAS_2}a). Finally, CPA values for T850 and $\Psi$700 are both characterised by a marked north-south division around 12$^{\circ}$N (Figure~\ref{fig:CPA_JAS}e,f).  The patterns indicate that in the north, high rainfall amounts are accompanied by lower T850, likely indicating a northward progression of the moist and cool monsoon layer, while in the south warm air at 850 hPa may indicate more instability on the following day.  With respect to $\Psi$700 low values in the north indicate that rainfall is accompanied by more cyclonic conditions, likely due to the trough passage of AEWs, while in the south weak anticyclonic conditions prevail.  

Most meteorological variables from Table~\ref{tab:era5_variables} show spatial patterns akin to those in Figure~\ref{fig:CPA_JAS}, though some feature hard to interpret local signals that entail a wider range of CPA values (e.g., 2T and SHR, see Appendix \ref{app:figures}, Figure~\ref{fig:CPA_JAS_2}b,d).  It is also worth mentioning that corresponding spatial structures for AUC largely agree with CPA (not shown).  Comparing JAS with the other four seasons, we find a high consistency in the discussed patterns that largely shift north- and southward with the seasonal evolution of the West African monsoon system (not shown).        

\begin{figure}[tb]
\centering
\includegraphics[width = 0.575 \textwidth]{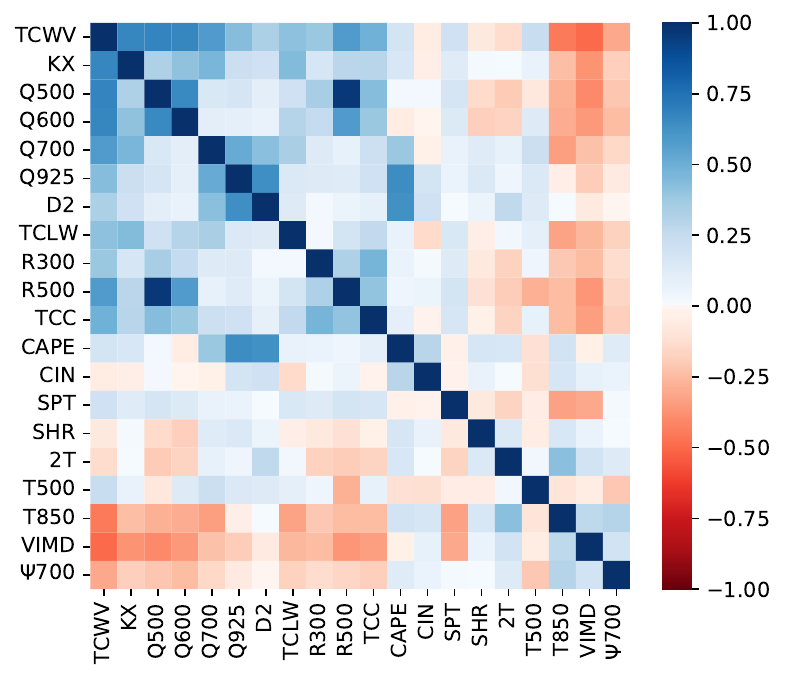}
\caption{Spatially averaged Spearman's rank correlation coefficient in season JAS between the ERA5 variables from Table \ref{tab:era5_variables}.  \label{fig:spear_mat_jas}} 
\end{figure}

For the construction of statistical models, correlations between predictor variables matter, as they hinder interpretation and may yield unstable statistical parameters.  Figure \ref{fig:spear_mat_jas} visualizes Spearman's rank correlation coefficients in JAS for the 20 predictor variables from Table~\ref{tab:era5_variables}.  We compute Spearman's coefficient at each grid point, and then average over grid points.  Note that here, we want the correlation coefficient to be symmetric (in contrast to the asymmetric relation between target and predictor variables).  This analysis has been conducted for all five seasons (Appendix \ref{app:figures}, Figure~\ref{fig:spear_mat}), but due to the large similarities between them, we discuss the peak monsoon season JAS only. 

Not surprisingly, there are generally high correlations between the moisture variables (TCWV, Q500, Q600, Q700, Q925, 2D, TCLW, R500, R300, and TCC), where it is noteworthy that R500 is more strongly correlated to Q500 than T500.  KX and CAPE show considerably different patterns, with KX being highly correlated with the moisture variables but surprisingly also associated with cold T850, which to some extent counteracts the impact of moister conditions.  CAPE is most sensitive to low-level moisture and associated with warm T850, as does CIN but to a smaller degree.  A positive SPT is weakly associated with a moister and warmer atmosphere, consistent with the southerly flow behind an AEW trough, where the moister atmosphere suppresses longwave cooling.  SHR, 2T, and T500 show overall weak and unsystematic correlations, in agreement with the difficult to interpret spatial patterns for CPA discussed above.  Finally, T850, VIMD, and $\Psi$700 are consistently negatively correlated with the moisture variables and KX, with the exception of 2D.  While the relation to VIMD is straightforward, T850 may indicate north-south movements of the monsoon layer, bringing overall moister or drier conditions.  The negative correlation between moisture variables and $\Psi$700 reflects the wet conditions associated with cyclonic disturbances, e.g., AEW troughs or vortices.

\section{Physics-based and data-driven forecast methods}  \label{sec:fct_model}

Forecasts for precipitation occurrence and precipitation amount ought to be probabilistic to account for the chaotic nature of the atmosphere, thus for the former they should output a probability of precipitation (PoP) and for the latter a probability distribution.  We investigate forecasts for precipitation occurrence and precipitation amount separately, which allows to connect our results to \citet{Vogel.2021} and \citet{Athul2023}, where the binary setting was considered only.  Furthermore, we can compare between the comparably easy task of producing PoP forecasts and the more challenging task of constructing probabilistic forecasts for precipitation amount.  To assess the skill of statistical and machine learning models it is essential to use baseline models to which to compare the forecast performance.  In the following subsections, different types of forecasting models are presented that are physics-based NWP models, purely data-driven statistical or machine learning techniques, or mixtures of both.  Table \ref{tab:fct} provides an overview of all considered approaches.

\renewcommand{\arraystretch}{1.1}

\begin{table}[t]
\caption{Overview of probabilistic forecast methods for precipitation accumulation, including general type, brief description, acronym, and availability of training data.  Methods marked with an asterisk$^*$ yield PoP forecasts only; for methods marked$^{**}$ we do not present results for PoP forecasts.  The final column notes from which year and month onward training data are available and used.  See text for details.  \label{tab:fct}}
\footnotesize
\centering
\begin{tabular}{llll} 
\toprule
Type             & Description                                         & Acronym         & Training \\
\toprule
Climatological   & monthly probabilistic climatology                   & MPC             & 2000 12 \\
\midrule
Physics-based    & ECMWF ensemble prediction system                    & EPS             & NA \\
                 & isotonic regression applied to EPS                  & EPS+ISO$^*$     & 2006 12 \\
                 & EMOS applied to EPS                                 & EPS+EMOS$^{**}$ & 2006 12 \\
                 & EasyUQ applied to HRES                              & HRES+EasyUQ     & 2001 12 \\
\midrule
Statistical      & logistic regresssion, baseline (5 predictors)       & Logit-base$^*$  & 2000 12 \\
                 & same, full model (25 predictors)                    & Logit-full$^*$  & 2000 12\\
                 & distributional index model, baseline (5 predictors) & DIM-base$^{**}$ & 2000 12 \\
                 & same, full model (25 predictors)                    & DIM-full$^{**}$ & 2000 12 \\
\midrule
Machine learning & EasyUQ applied to convolutional neural network      & CNN+EasyUQ      & 2000 12 \\
\midrule
Hybrid           & mixture of HRES+EasyUQ and CNN+EasyUQ               & Hybrid          & NA \\ 
\bottomrule
\end{tabular}
\end{table}

As discussed in section \ref{sec:data_methods}, our evaluation period for 24-hour forecasts of precipitation amount and precipitation occurrence ranges from 1 December 2010 to 30 November 2019.  The DJF season runs across two subsequent calendar years and we generally assign it to the second year.  Reporting a yearly seasonal or overall mean instead of a single mean score over the complete evaluation period allows for a more distinct comparison between forecasting models and provides insights into the temporal evolution of forecast skill.  

Except for the ECMWF ensemble prediction system (EPS), all types of forecasting methods require training data and some form of training procedure.\footnote{Our Hybrid model combines the CNN+EasyUQ and HRES+EasyUQ forecasts in a way that does not require additional training.}  In this study, we use annually growing, expanding training sets that resemble operational settings, where only past data are available.\footnote{Nonparametric statistical methods such as IDR and especially machine learning approaches benefit from having as much (relevant) training data available as possible.  Subject to this caveat, the predictive performance generally does not depend very much on the details of the training scheme.  For example, we also implemented the EPS+EMOS technique using a rolling training period of the most recent 730 days and obtained similar results.}  The initial training period ranges from the first day of the month in the right most column of Table \ref{tab:fct} (hereinafter, the start date) to 30 November 2010, and the thus trained methods are used to generate day-ahead 24-hour forecasts for the period from 1 December 2010 to 30 November 2011.  Then, we successively add one more year to the training period, ranging now from the start date through 30 November in year $2010 + x$, and use the thus trained methods to generate forecasts for the 12-month period that begins on 1 December in year $2010 + x$, where $x \in  \{ 1, \dots, 8 \}$.  This procedure is followed until training is on data through 30 November 2018 and the thus trained methods are used to generate forecasts for 1 December 2018 through 30 November 2019.  Thus, there are nine evaluation folds in total, which we associate with calendar years 2011, \dots, 2019, respectively.

\subsection{Climatological forecasts}  \label{sec:climatological}

Arguably, the simplest possible type of probabilistic forecast is a climatology constructed from past observations.  Here we use GPM IMERG to construct a monthly probabilistic climatology (MPC).  The MPC forecast for a specific valid date is an ensemble constructed by using all past observations from the month at hand.  For example, for a test date in January 2014, the MPC forecast is constructed based on data from January 2001 to January 2013, which yields an ensemble of size $31 \times 13 = 403$.  To obtain the MPC PoP forecast, the relative frequency of ensemble members with rainfall exceeding 0.2 mm is computed.  

\subsection{Physics-based forecasts}  \label{sec:physics-based}

Our comparison includes raw and postprocessed probabilistic forecasts from physics-based numerical weather prediction (NWP) models run by the ECMWF (section \ref{sec:ECMWF}).  The postprocessed forecasts require training, for which we use expanding training sets with start dates listed in Table \ref{tab:fct} as described above.  Training is performed at each grid point individually.

\subsubsection{Operational ECMWF NWP ensemble}  \label{sec:EPS}

The operational ECMWF ensemble prediction system (EPS) comprises 51 NWP runs, namely, a control member and 50 perturbed members.  Just as for the climatological MPC approach, the EPS PoP forecast is the relative frequency of members that exceed 0.2 mm.

\subsubsection{Statistically postprocessed ECMWF NWP ensemble}  \label{sec:postprocessed}

Statistical postprocessing is used to correct for systematic biases in raw ensemble forecasts.  Here we use Ensemble Model Output Statistics (EMOS), originally developed by \citet{Gneiting2005}, to generate full predictive probability distributions by linking ensemble information to distributional parameters.  The optimal coefficients are found by optimizing a performance metric on training data.  

In the binary case, we recalibrate the EPS PoP by using nonparametric isotonic regression \citep{zadrozny2002}, here referred to as EPS+ISO.  For precipitation amount, we apply the EMOS technique proposed by \citet{Scheuerer2014} which models positive rainfall accumulations with generalized extreme value distributions, to generate the EPS+EMOS forecast.  While EPS+EMOS induces a PoP forecast, the predictive performance is very similar, though typically slightly inferior, to EPS+ISO.  Therefore, we do not report results for the respective PoP forecasts (cf.~Table \ref{tab:fct}).

\subsubsection{EasyUQ on the HRES model}  \label{sec:HRES}

The high resolution (HRES) model from ECMWF generates a deterministic NWP forecast.  We use the EasyUQ technique, introduced in section \ref{sec:EasyUQ}, to transform this single-valued forecast into a postprocessed predictive distribution, to yield the HRES+EasyUQ forecast.

\subsection{Statistical forecasts}  \label{sec:statistical}

Statistical approaches use training data to learn relationships between a target variable and one or more predictor variables.  Here, the target variable is precipitation amount at a given grid point, which in the case of precipitation occurrence is thresholded at 0.2 mm.  We use logistic regression to obtain PoP forecasts and Distributional single Index Models \citep[DIMs;][]{DIM} for probabilistic forecasts of precipitation amount, based on predictor variables from section \ref{sec:predictors}.  Statistical models require training, and we use annually expanding training sets with start date in December 2000 (Table \ref{tab:fct}) as described above.  Training is performed at each grid point individually.

The analysis in section \ref{sec:predictors} provides a thorough understanding of the influence of the selected variables from Table \ref{tab:era5_variables} on precipitation occurrence and amount, and enables to link them to typical seasonal weather phenomena.  However, overall, the effect of meteorological variables on precipitation is similar across seasons when taking into account the latitudinal shifts associated with the monsoon system.  As a consequence, we found little difference in model performance between fitting models on seasonal data versus the whole available training period, as temporal effects such as seasonal changes can be captured by predictor variables that encode the day of the year. Therefore, instead of fitting seasonal models, we train models that apply year-round.  

We distinguish baseline models with two predictors that encode the day of the year and three correlated rainfall predictors (section \ref{sec:corrs}) from full models that additionally use 20 predictor variables from ERA5 (section \ref{sec:preds_era5}).  To prevent a statistical model from overfitting, regularization techniques can be applied.  However, in this experiment the performance of the statistical models, which use modest numbers of at most 25 predictor variables only, does not improve when using the regularization techniques we tested.  Consequently, we refrain from performing any feature selection beyond the choices made in section \ref{sec:predictors}, which were driven by meteorological expertise and extant literature in atmospheric physics.

\subsubsection{Logistic regression}  \label{sec:Logit}

We use logistic regression (Logit) models to generate statistical PoP forecasts.  Specifically, let $m$ be the number of predictor variables, which we denote by $x_1, \ldots, x_m$, and let $p$ be the PoP forecast.  The logistic regression model then is of the form
\begin{align}  \label{eq:logit}
\text{logit}(p) = \log \frac{p}{1-p} = \alpha_0 + \sum_{j=1}^m \alpha_j x_j, 
\end{align}
where the statistical coefficients $\alpha_0, \alpha_1, \dots, \alpha_m$ are estimated from training data.  Our baseline model (Logit-base) originates from \citet{Vogel.2021} and \citet{Athul2023} and uses $m = 5$ predictor variables, namely, three correlated rainfall predictors $x_1, x_2$, and $x_3$ at temporal lags of one, two, and three days, respectively, as described in section \ref{sec:corrs}, and two variables $x_4 = \sin(2\pi d / 365)$ and $x_5 = \cos(2\pi d / 365)$ that depend solely on the day of the year $d$.  The full model (Logit-full) extends to $m = 25$ predictor variables in eq.~\eqref{eq:logit}, now including the twenty ERA5 variables from Table \ref{tab:era5_variables}.

\subsubsection{Distributional index models}  \label{sec:DIM}

To produce probabilistic forecasts for accumulated precipitation we use the Distributional (single) Index Model (DIM) approach introduced by \citet{DIM}, which combines the classical single index model with Isotonic Distributional Regression \citep[IDR;][]{Henzi.2021}.  In a nutshell, an index is learned that represents the conditional mean of the target variable, here log-transformed precipitation accumulation, and then a predictive distribution is estimated nonparametrically under a stochastic ordering constraint.  As before, let $x_1, \ldots, x_m$ be predictor variables, and let $y$ be the target, namely, precipitation accumulation.  The index model assumes the relationship 
\begin{align}  \label{eq:DIM}
\log \left( y + \frac{1}{100} \right) =  \beta_0 + \sum_{j=1}^m \beta_j x_j,  
\end{align}
where the statistical coefficients $\beta_0, \beta_1, \dots, \beta_m$ are learned from training data.  Subsequent to the training of the index model, the nonparametric IDR distributions are estimated on the same training data, augmented with the fitted index values.  We distinguish a baseline model (DIM-base, $m = 5$) and an extended model (DIM-full, $m = 25$ in eq.~\eqref{eq:DIM}), for which we use the same sets of predictor variables as in the Logit approach from section \ref{sec:Logit}.

Note that PoP forecasts can be extracted from the DIM-base and DIM-full distributions.  These yield similar, though slightly inferior, results than the Logit-base and Logit-full PoP forecasts, respectively, and so we do not report the respective scores (cf.~Table \ref{tab:fct}).

\subsection{Machine learning based forecasts: CNN+EasyUQ}  \label{sec:CNN}

The aforementioned statistical models are applied at each grid point individually.  Thus, including spatial information has to be done by manually engineering features accordingly, such as the correlated rainfall predictors in section \ref{sec:corrs}.  In contrast, Convolutional Neural Network (CNN) models operate directly on the two-dimensional input space and can learn spatial relations from the data without the need to extract spatial information beforehand.  CNN models are most commonly used for image tasks, where the input usually is a two- or three-dimensional array of pixel values.  The gridded weather data over our evaluation domain can be envisaged as two-dimensional pseudo images of size $19 \times 61$.  These dimensions correspond to latitude and longitude, respectively, spanning the study domain (Figure~\ref{fig:area}) from $0^\circ$ to $18^\circ$ N and $25^\circ$ W to $35^\circ$ E, respectively, with a grid resolution of $1^\circ \times 1^\circ$.  With a suitable architecture, a single CNN model produces a two-dimensional array with forecasts for all grid points at once, instead of training models at each grid point individually.  Due to their inherent inductive bias towards local neighborhood connectivity, CNNs are well-suited for predicting precipitation on the $19 \times 61$ grids, as they effectively exploit spatial correlations and structures within a grid, recognizing patterns within local areas that may be indicative of specific weather conditions.  For this reason, the three correlated rainfall predictors from section \ref{sec:corrs} are replaced by $19 \times 61$ grids of IMERG precipitation accumulations (section \ref{sec:IMERG}) at temporal lags of one, two, and three days, respectively. 

Motivated by their successful application in related meteorological tasks \citep{Ayzel2020, Weyn2020, Lagerquist2021, Chapman2022, Otero2023}, we employ a CNN architecture in the form of the U-Net \citep{Ronneberger2015UNet}.  The architecture of the U-Net consists of a contracting (downsampling) path and an expansive (upsampling) path, which are symmetric in terms of individual layer properties, giving it a U-like shape.  We make use of max pooling operations for downsampling and transposed convolutions for upsampling layers.  A crucial feature of the U-Net is skip connections between layers of the same size in the contracting and expanding paths.  Applied to the precipitation data grid, these connections allow the network to use information from multiple resolutions, combining the context from the contracting path with the localization information from the expansive path.  This allows to model longer spatial range dependencies in the data.  To avoid overfitting, we also make use of Dropout \citep{Srivastava2014dropout} throughout the network architecture.

To transform the deterministic precipitation forecasts of the CNN model into probabilistic forecasts, the EasyUQ technique introduced in Section \ref{sec:EasyUQ} is applied at each grid point individually, subsequent to the training of the index model, and based on the same training data as for the neural network, augmented with the deterministic CNN output.  As noted, the resulting CNN+EasyUQ forecast distributions are discrete and have mass exclusively at outcomes observed during training.  Code for the implementation of the CNN+EasyUQ approach in Python \citep{Python} is publicly available under \url{https://github.com/evwalz/precipitation}.  Once more we emphasize that, while our usage of EasyUQ in concert with the CNN model is novel, we employ standard choices, such as quadratic loss and 3x3 convolutional kernels, for the neural network architecture and neural network training. 

\subsection{Hybrid approaches}  \label{sec:Hybrid}

NWP models represent the physical laws of atmospheric dynamics through a set of differential equations.  Statistical or machine learning based approaches, on the other hand, do not encode physical laws but learn patterns based exclusively on past data.  A hybrid model is a combination of both approaches and thus can benefit from both the physical expertise embodied in NWP output and the flexibility of data-driven approaches.  In this paper, we base hybrid approaches on the deterministic HRES forecast from section \ref{sec:HRES} and the deterministic CNN forecast from section \ref{sec:CNN}.  We consider three approaches to obtain probabilistic forecasts from the deterministic HRES and CNN forecasts.  First, the NWP forecast can be used as an additional gridded feature in the CNN model, followed by grid point based application of EasyUQ.  Secondly, we can apply IDR using both deterministic forecasts as input features.  Lastly, a simple approach is to use a weighted or unweighted average of the predictive distributions generated by HRES+EasyUQ and CNN+EasyUQ.  We found experimentally that the first two approaches do not improve predictive ability, generally showing similar forecast performance than the CNN+EasyUQ forecast.  The last approach in its most basic form of an equal average between the HRES+EasyUQ and CNN+EasyUQ distributions, which does not require any additional training, shows slight forecast improvements.  It is therefore selected and referred to as the Hybrid model.

\section{Forecast evaluation}  \label{sec:evaluation}

In this section we report major findings from the forecasting experiment.  The discussion concentrates on the peak monsoon season JAS, but we provide results for the other seasons in additional figures in Appendix \ref{app:figures}.  As described at the start of section \ref{sec:fct_model}, our experiment uses expanding training sets to learn the forecasting models, and we frequently report annual results from the evaluation folds for 2011, \dots, 2019.  As evaluation metrics, the mean Brier score (BS) from eq.~\eqref{eq:BS} and the mean continuous ranked probability score (CRPS) from eq.~\eqref{eq:CRPS} are used.

\subsection{Effects of variable selection in statistical models}  \label{sec:effects}

\begin{figure}[t]
\includegraphics[width = \textwidth]{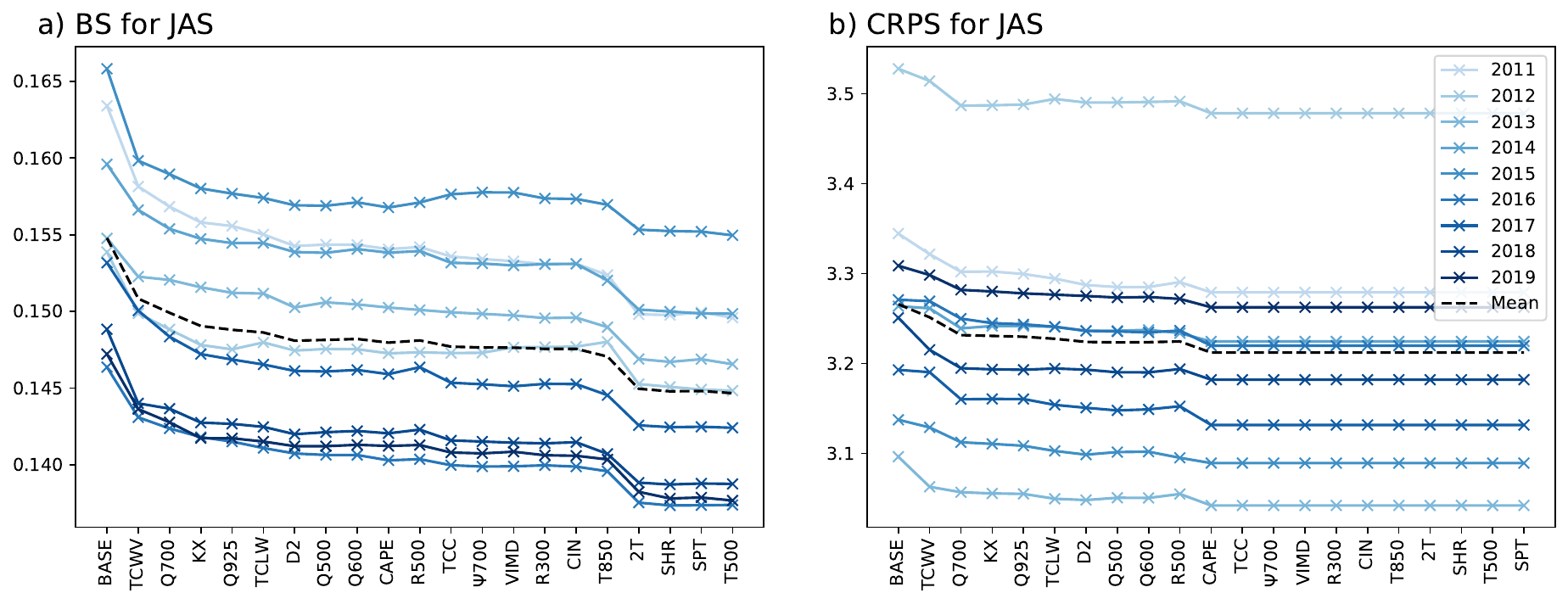}
\caption{a) Mean Brier score (BS) for the Logit PoP forecast under successive addition of the predictor variables displayed on the horizontal axis.  The base model includes three correlated rainfall predictors and two time features.  The BS is averaged over space and season JAS for evaluation folds from 2011 to 2019.  b) Corresponding display for the DIM forecast of precipitation accumulation and the CRPS.  \label{fig:feature_sel}}
\end{figure}

To better understand the influence of the number of predictor variables on the forecast performance of the statistical models, namely, the Logit PoP forecast (section \ref{sec:Logit}) and the DIM forecast for precipitation accumulation (section \ref{sec:DIM}), a visual analysis is provided in Figure~\ref{fig:feature_sel}.  Starting with the mean score of the base model, which has five predictor variables, one more variable is successively added and the corresponding mean score is shown, until the full model with 25 predictor variables is reached.  The variables are selected in the order of the distance between 0.5 and the mean AUC respectively CPA computed without splitting into seasons.\footnote{An AUC or CPA value of 0.5 suggests a useless feature.}

Figure \ref{fig:feature_sel}a shows that the addition of TCWV to the Logit base model yields an improvement of the BS on the order of 5\% in all years.  Small further improvements of less than 1\% are obtained by adding mid-level humidity (Q700) and static stability (KX).  The addition of further variables yields minor improvement only, with the exception of 2m temperature (2T), which leads to an improvement comparable to Q700 and KX, despite AUC values barely above 0.5 (Figure~\ref{fig:AUC}d).  Qualitatively, improvements in CRPS per predictor regarding precipitation amount (Figure~\ref{fig:feature_sel}b) show similar results, yet the percentage improvements are smaller such that adding variables other than TCWV and Q700 barely improves performance.  Generally, the performance difference between years is large, and the ranking of the years differs between the BS, where the lowest values are seen for 2017, and the CRPS, where they are seen for 2013.  Results for seasons other than JAS are qualitatively similar, except that the overall level of the scores varies strongly between seasons (Appendix \ref{app:figures}, Figures~\ref{fig:feature_sel_} and \ref{fig:feature_sel__}). 

\subsection{Comparative evaluation of predictive performance}  \label{sec:comparative}

\begin{figure}[t]
\includegraphics[width = \textwidth]{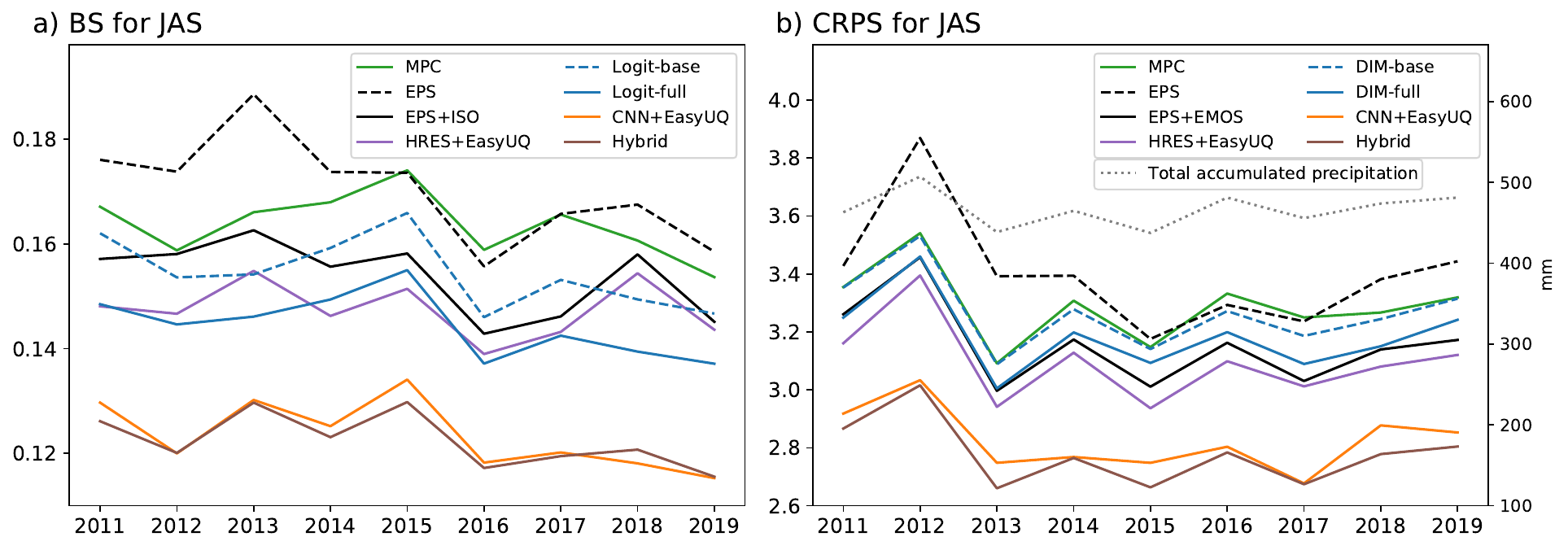}
\caption{a) Mean Brier score (BS) for PoP forecasts from Table \ref{tab:fct}, averaged over space and season JAS for evaluation folds from 2011 to 2019.  b) Corresponding display for the continuous ranked probability score (CRPS) and probabilistic forecasts of precipitation amount, along with spatially averaged total accumulated precipitation for JAS, in the unit of millimeters.  \label{fig:yearly_bs_crps}}
\end{figure}

Figure \ref{fig:yearly_bs_crps}a visualizes the mean Brier score (BS) in season JAS for the PoP forecasting models from Table \ref{tab:fct}.  Similar to the results in \citet{Vogel.2021}, the ECMWF ensemble prediction system (EPS) shows inferior or, in later years, comparable performance than MPC, and both EPS and MPC are outperformed by a simple logistic regression approach based on correlated rainfall predictors only (Logit-base).  The inclusion of ERA5 predictors into the logistic regression model (Logit-full) leads to a clear improvement beyond the post-processed EPS-ISO PoP forecast.  Surprisingly, the HRES+EasyUQ forecast shows better performance than the ensemble-based EPS+ISO forecast.  The CNN+EasyUQ PoP forecast outperforms all other methods, except for the Hybrid forecast, which shows nearly the same performance.

The mean CRPS for the forecasting models for precipitation accumulation from Table \ref{tab:fct} is displayed in Figure~\ref{fig:yearly_bs_crps}b.  Through 2014, EPS clearly shows the lowest forecast skill; thereafter, its skill improves and gets close to the performance of MPC and DIM-base.  Unlike the Logit-full PoP forecast, DIM-full does not outperform the post-processed EPS+EMOS forecast.  The HRES+EasyUQ approach yields better scores than EPS+EMOS, probably due to the flexibility of the EasyUQ forecast distributions.  The CNN+EasyUQ approach shows a forecast improvement within the evaluation period, and the Hybrid model performs similar or slightly better for some years.  As can be seen by the dotted line giving the JAS area-averaged rainfall, the mean CRPS co-varies with the total rainfall amount, thus the years with the best performance are usually also the driest. 

Figures \ref{fig:seasonal_bs_mean} and \ref{fig:seasonal_crps_mean} in Appendix \ref{app:figures} show analogous evaluation results for all five seasons.  Throughout, the CNN+EasyUQ and Hybrid forecasts perform similarly to each other, and outperform their competitors by considerable margins.  

\subsection{Spatial structure of predictive performance}  \label{sec:spatial}

\begin{figure}[t]
\centering
\includegraphics[width = \textwidth]{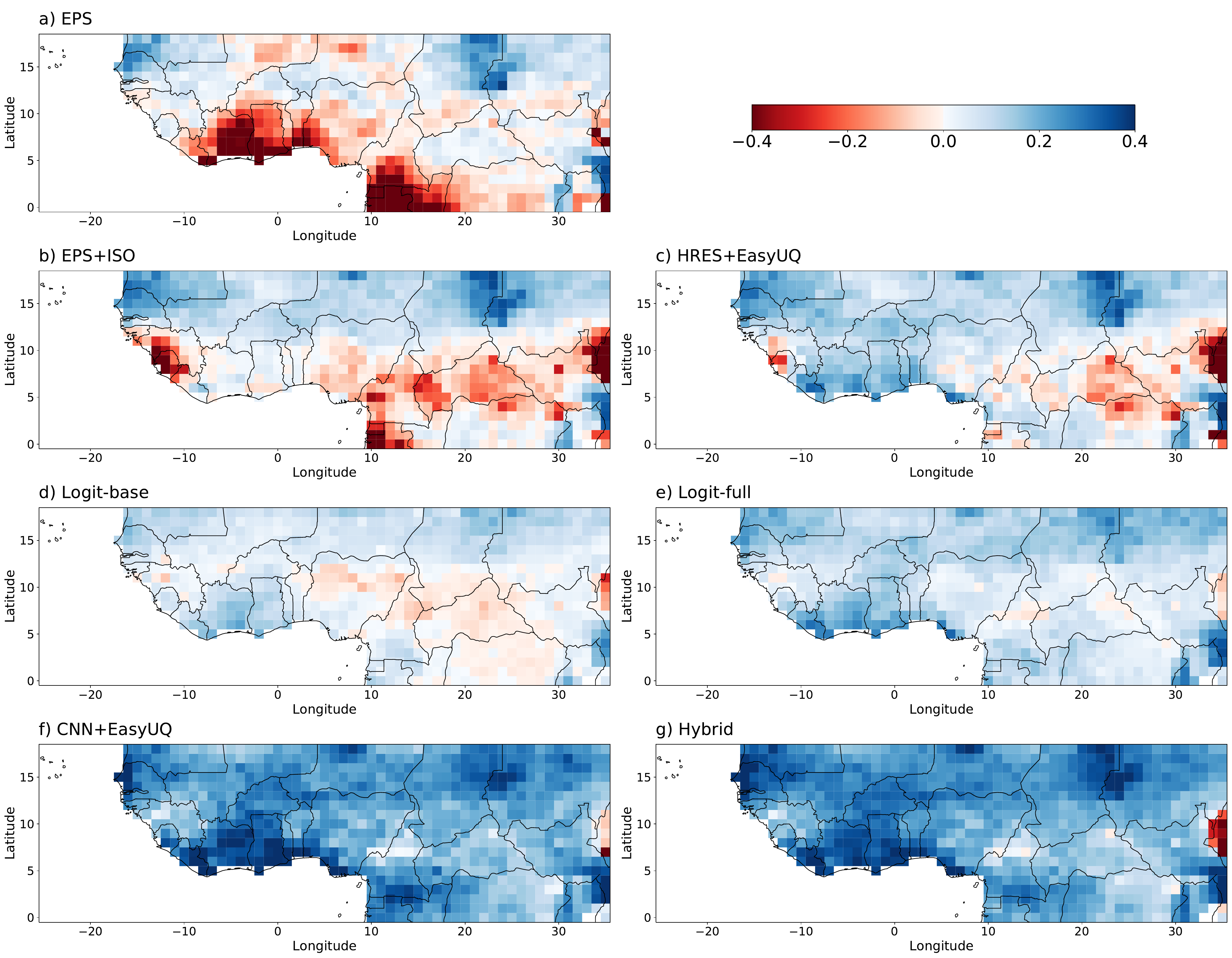}
\caption{Spatial structure of the Brier skill score for probability forecasts of precipitation occurrence with a) EPS, b) EPS+ISO, c) HRES+EasyUQ, d) Logit-base, e) Logit-full, f) CNN+EasyUQ, and g) the Hybrid forecast from Table \ref{tab:fct}, relative to MPC as baseline, for season JAS and combined evaluation folds from 2011 to 2019.  \label{fig:skillscore_bs}}
\end{figure}

\begin{figure}[t]
\centering
\includegraphics[width = 0.98 \textwidth]{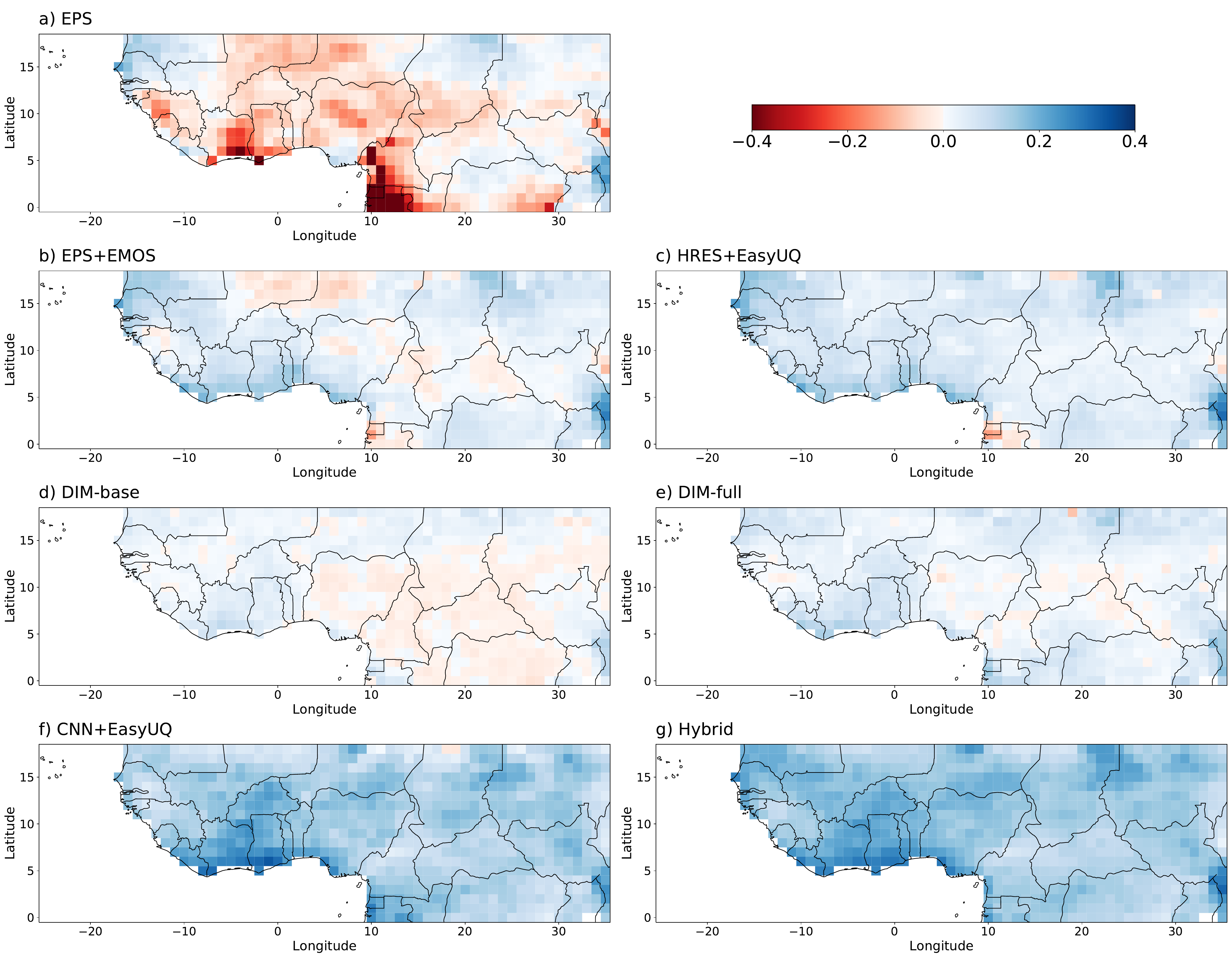}
\caption{Spatial structure of the CRPS skill score for probabilistic forecasts of precipitation accumulation with a) EPS, b) EPS+EMOS, c) HRES+EasyUQ, d) DIM-base, e) DIM-full, f) CNN+EasyUQ, and g) the Hybrid forecast from Table \ref{tab:fct}, relative to MPC as baseline, for season JAS and combined evaluation folds from 2011 to 2019.  \label{fig:skillscore_crps}}
\end{figure}

For an understanding of spatial patterns of forecast performance, skill score plots of the forecast approaches considered here with MPC as reference forecast are shown in Figure~\ref{fig:skillscore_bs} for precipitation occurrence and in Figure~\ref{fig:skillscore_crps} for precipitation accumulation, both for the JAS peak monsoon season and across evaluation folds.

With respect to the PoP forecasts for rainfall occurrence, EPS shows negative skill relative to MPC over the southern parts of the study domain, particularly over the relatively dry areas along the Guinea coast, over Gabon and southern Cameroon, where rainfall tends to be rather localized and short-lived such that precipitation occurrence is hard to predict (Figure~\ref{fig:skillscore_bs}a).  Senegal/Mauritania and Chad/Sudan are the only areas with considerable positive skill, while the rest of the domain ranges close to zero.  Applying statistical postprocessing (EPS+ISO, Figure~\ref{fig:skillscore_bs}b) removes the large negative skill along the Guinea Coast but shows remaining issues in a stretch from Nigeria to South Sudan with mostly weakly negative skill.  Remarkably, postprocessing deteriorates skill around the highlands in Guinea/Sierra Leone and westernmost Ethiopia.  Over the Sahel, in contrast, the postprocessing leads to an overall improvement and consistently positive skill.  A possible reason is the stronger influence of predictable features such as AEWs or midlatitude perturbations here in contrast to the more stochastic rains in the south \citep{Athul2023}.  The comparison between the EPS+ISO and HRES+EasyUQ (Figure~\ref{fig:skillscore_bs}c) demonstrates that for forecasts at individual sites there is no added value in running an NWP ensemble system, even after postprocessing.  The structures are fairly consistent (e.g., with problematic regions in Guinea/Sierra Leone, the Central African Republic, South Sudan, and Ethiopia) but the values are consistently more positive for the HRES+EasyUQ technique, which is based on HRES model alone, as opposed to using an ensemble. 

Moving to the data-based approaches (Figure~\ref{fig:skillscore_bs}d–-g) we see consistent improvement over most areas of the study domain, though PoP forecasts for western Ethiopia remain a challenge, possibly related to the rough topography in this area.  While in the simpler Logit-base approach (Figure~\ref{fig:skillscore_bs}d) some areas of negative skill remain, the inclusion of additional predictors in Logit-full (Figure~\ref{fig:skillscore_bs}e) leads to a consistent improvement and thus positive skill almost everywhere in the study region.  It is also noteworthy that the Logit models generate overall smoother skill fields compared to the physics-based approaches.  Finally, the CNN+EasyUQ and Hybrid methods (Figure~\ref{fig:skillscore_bs}f,g) outperform all other approaches to a large extent, reaching up to 40\% improvement relative to the climatological benchmark MPC.  The improvement relative to EPS is particular impressive over the Guinea coastal region (e.g., Ivory Coast and Ghana), where EPS performs much worse than MPC, and illustrates the ability of the CNN to learn complex physical relationships that determine local rainfall probability.  The inclusion of NWP information from the HRES model in the Hybrid approach yields small improvements in some places but no clear advance relative to CNN+EasyUQ.  This demonstrates that knowing the ambient conditions shortly before the beginning of the 24-hour forecast period is much more important than knowledge of the forecast evolution during that period.

The corresponding analysis for rainfall amount (Figure~\ref{fig:skillscore_crps}) reveals many parallels to rainfall probability.  EPS (Figure~\ref{fig:skillscore_crps}a) stands out as having many areas of negative CRPS skill, with an overall similar structure to the occurrence analysis (Figure~\ref{fig:skillscore_bs}a).  Postprocessing (EPS+EMOS, Figure~\ref{fig:skillscore_crps}b) cures many issues of EPS, leading to mostly weakly positive skill, but does not perform as well as the computationally much less expensive HRES+EasyUQ technique (Figure~\ref{fig:skillscore_crps}c).  Here the skill fields for amount are overall smoother than for occurrence with less contrast between the Sahel and the southern areas.  The DIM models (replacing the Logit models for amount) show negligible further advance.  The skill of DIM-base (Figure~\ref{fig:skillscore_crps}d) is close to zero everywhere with a negative area in the southeast and positive elsewhere, while the inclusion of additional predictors (DIM-full, Figure~\ref{fig:skillscore_crps}e) slightly improves skill over most areas.  Finally, as for occurrence, the machine learning based CNN+EasyUQ and Hybrid methods (Figure~\ref{fig:skillscore_crps}f,g) outperform all other approaches to a large extent with positive CRPS skill of up to 30\%.  Here the Hybrid approach leads to a more considerable improvement relative to CNN+EasyUQ, yielding fairly equal skill improvement across the entire, quite heterogeneous domain.  These improvements are more prominent in areas where the physics-based HRES model may better represent the time evolution of dynamical features such as AEWs and extratropical influences.

\begin{figure}[h]
	\centering
	\includegraphics[width = \textwidth]{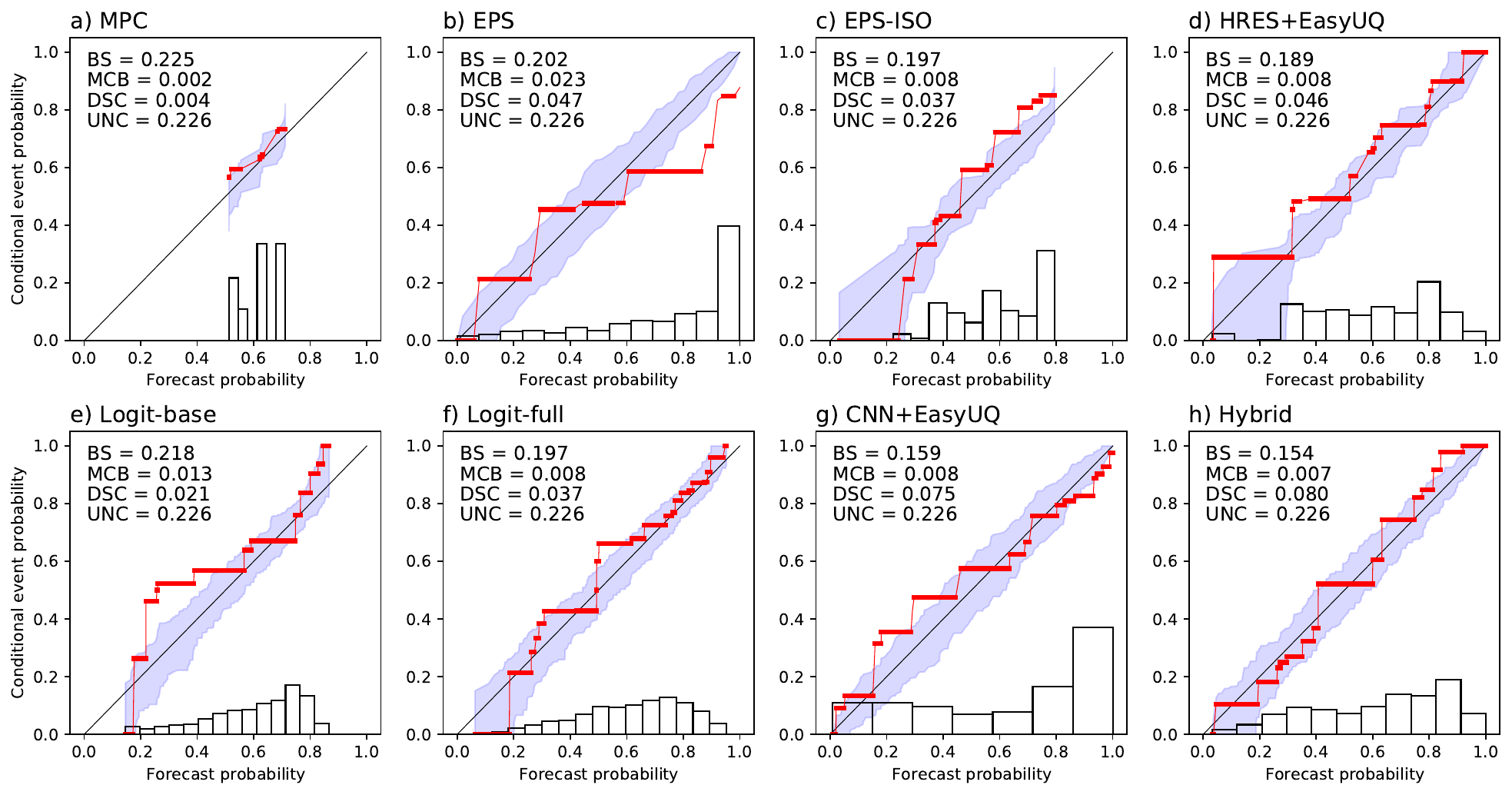}
	\caption{Reliability diagrams for PoP forecasts at the grid point closest to Niamey ($13^{\circ}$N, $2^{\circ}$E) with a) MPC, b) EPS, c) EPS+ISO, d) HRES+EasyUQ, e) Logit-base, f) Logit-full, g) CNN+EasyUQ, and h) the Hybrid approach from Table \ref{tab:fct}, for season JAS and combined evaluation folds from 2011 to 2019, with 90\% consistency bands under the assumption of calibration \citep{Dimitriadis2021}.  The panels also show the mean Brier score (BS) and its miscalibration (MCB), discrimination (DSC), and uncertainty (UNC) components from eq.~\eqref{eq:decomposition}.  The histograms along the horizontal axis show the distribution of the forecast probabilities.  \label{fig:corp}}
\end{figure}

\subsection{Calibration and discrimination ability}  \label{sec:calibration}

We now assess the calibration and discrimination ability of the forecasts.  Following \citet{Vogel.2021} and \citet{Athul2023}, reliability diagrams for the PoP forecasts from Table \ref{tab:fct} at the grid point closest to Niamey ($13^{\circ}$N, $2^{\circ}$E) are presented in Figure~\ref{fig:corp}.\footnote{Python code for computation and plotting \url{https://github.com/evwalz/corp_reldiag}.}  The panels use the CORP approach of \citet{Dimitriadis2021} and show the decomposition \eqref{sec:metrics} of the mean Brier score (BS) into miscalibration (MCB), discrimination (DSC), and uncertainty (UNC) components.  Instead of considering each evaluation fold separately, the decomposition is computed once on forecasts in the peak monsoon season JAS from all nine evaluation years together.  If the reliability curve is close to the diagonal, a PoP forecast is calibrated (reliable).  Deviations from the diagonal indicate some type of miscalibration: S-shaped curves indicate underconfidence (PoP too close to center), inverse S-shaped curves correspond to overconfidence (PoP too extreme), and curves that are below (above) the diagonal indicate biased PoP.  The climatological MPC PoP forecast has a very limited range of forecast values and lacks discrimination ability, but shows excellent calibration.  The poor calibration of the raw ENS PoP is corrected by post-processing (ENS+ISO).  In agreement with the findings in \citet{Vogel.2021} and \citet{Athul2023}, the Logit-base PoP forecast is well calibrated and has moderate discrimination ability.  In comparison, Logit-full shows a lower BS (more skillful PoP forecasts) reflected in both better calibration and improved discrimination ability.   The CNN+EasyUQ and Hybrid techniques show superior performance --- they are similarly well calibrated as EPS-ISO and Logit-full but show considerably higher discrimination ability.

\begin{figure}[h]
\includegraphics[width = \textwidth]{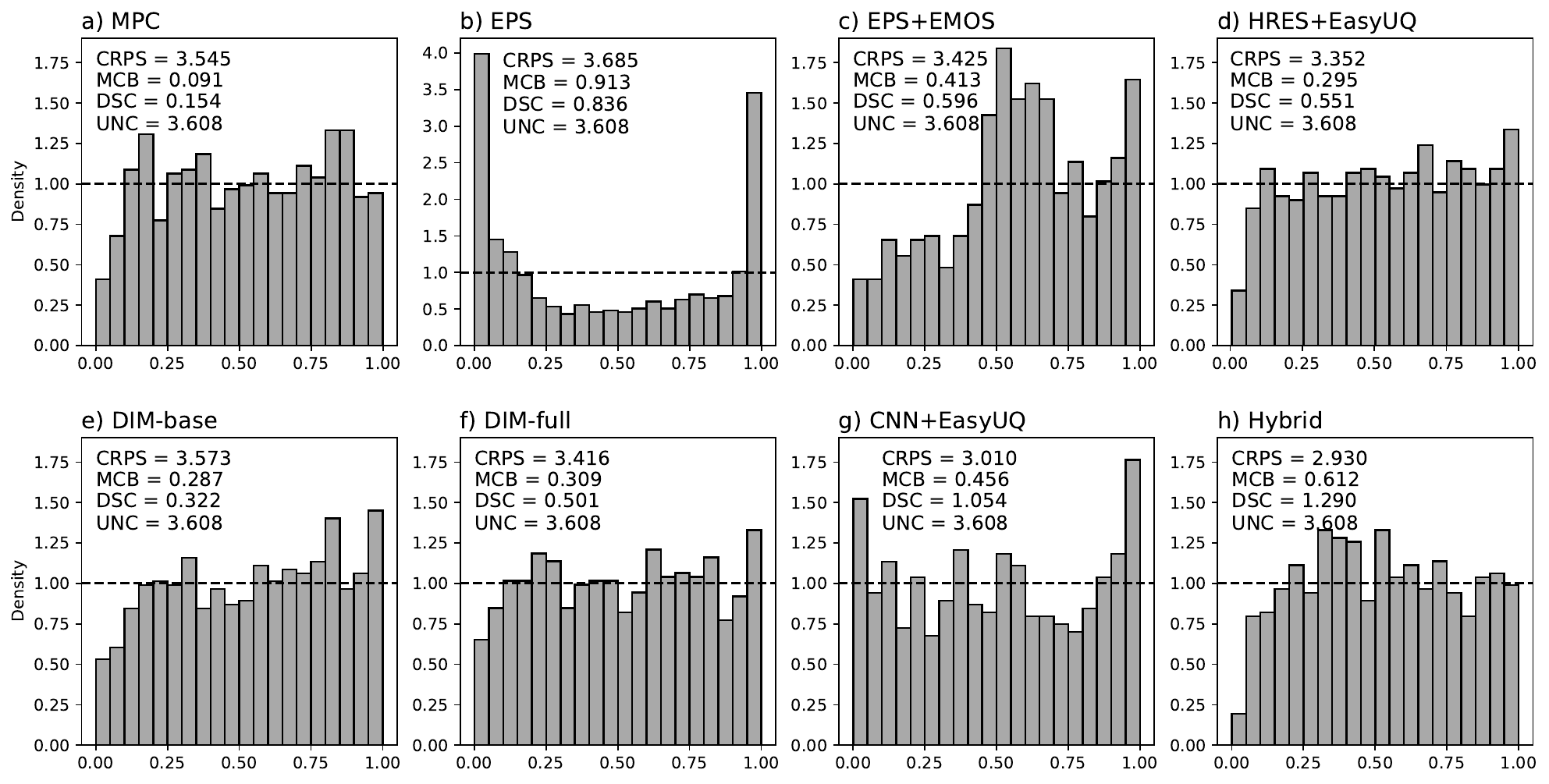}
\centering
\caption{Probability Integral Transform (PIT) histograms for probabilistic forecasts of precipitation accumulation at the grid point closest to Niamey ($13^{\circ}$N, $2^{\circ}$E) with a) MPC, b) EPS, c) EPS+EMOS, d) HRES+EasyUQ, e) DIM-base, f) DIM-full, g) CNN+EasyUQ, and h) the Hybrid approach from Table \ref{tab:fct}, for season JAS and combined evaluation folds from 2011 to 2019.   The panels also show the mean continuous ranked probability score (CRPS) and its miscalibration (MCB), discrimination (DSC), and uncertainty (UNC) components from eq.~\eqref{eq:decomposition}.  The vertical scale of the histograms is shared across forecasts, except for EPS.  \label{fig:pit}}
\end{figure}

To assess the calibration of the probabilistic forecasts for accumulated precipitation at the grid point closest to Niamey, Figure~\ref{fig:pit} shows Probability Integral Transform (PIT) histograms.  For the MPC and EPS ensemble forecast, a universal PIT (uPIT) histogram is shown \citep{Vogel.2018}; for the other methods, the randomized version of the PIT is used \citep[eq.~(1)]{Tilmann_Johannes_Calibration}.  A uniform histogram indicates calibrated forecasts while a U-shaped (hump-shaped) histogram suggests underdispersed (overdispersed) forecasts, meaning that the forecasts are overconfident (underconfident).  Skewed histograms indicate biases.  The ECMWF ensemble (EPS) is underdispersed, which is corrected for in the EPS+EMOS forecast, though a bias remains.  The other forecasts show PIT histograms that are nearly uniform.  The associated decomposition \eqref{eq:decomposition} of the mean CRPS demonstrates the superior calibration of the climatological MPC forecast and the outstanding discrimination ability and overall predictive performance of the CNN+EasyUQ and Hybrid approaches. 

\begin{figure}[t]
\includegraphics[width = 0.49 \textwidth]{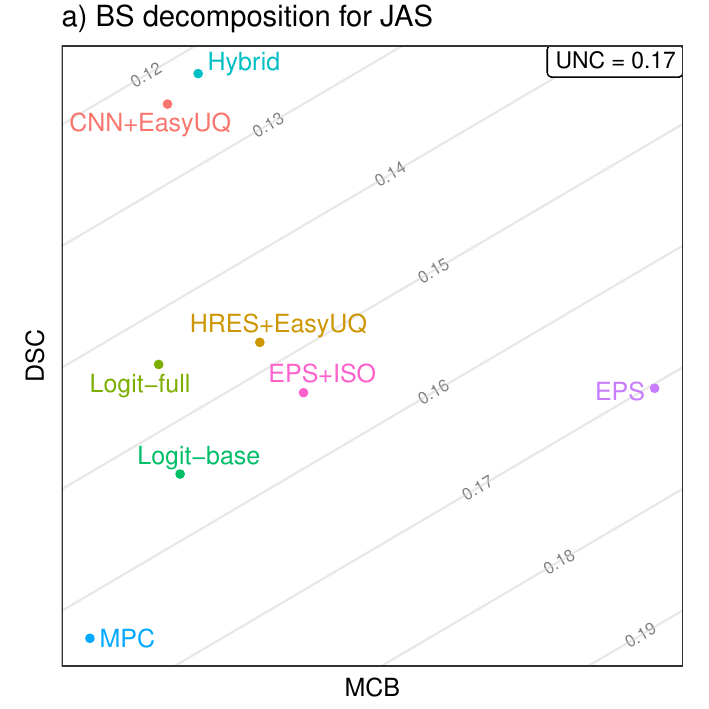}
\includegraphics[width = 0.49 \textwidth]{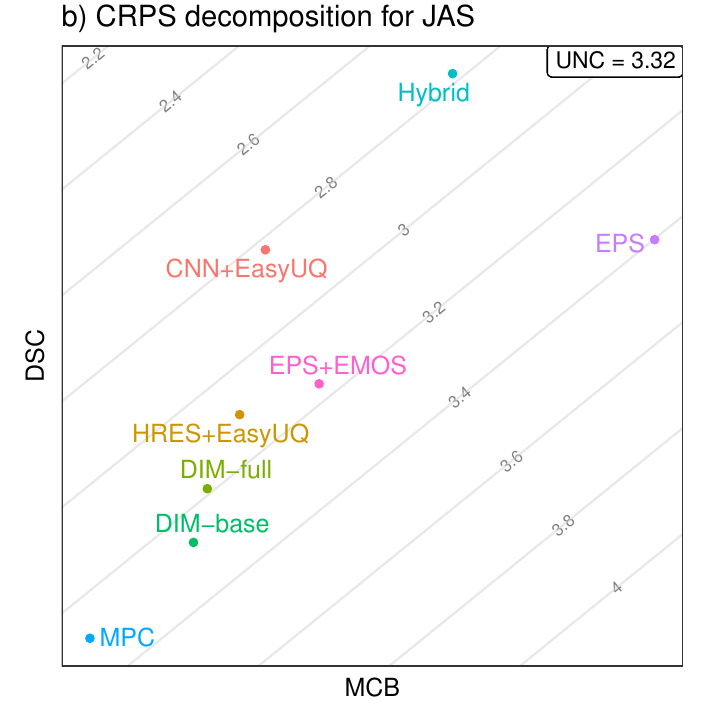}
\caption{Miscalibration (MCB), discrimination (DSC), and uncertainty (UNC) components of a) the mean Brier score (BS) for probability forecasts of precipitation occurrence, and b) the mean continuous ranked probability score (CRPS) for probabilistic forecasts of precipitation accumulation in the unit of millimeters, as described in Table \ref{tab:fct}.  The score decomposition in \eqref{eq:decomposition} is applied at each grid point, based on the combined evaluation folds from 2011 to 2019, and the mean score and score components are then averaged over grid points.  Parallel lines correspond to equal mean scores.  \label{fig:deco}}
\end{figure}

Finally, we use the decomposition of the mean Brier score (BS) or mean continuous ranked probability score (CRPS) into miscalibration (MCB), discrimination (DSC), and uncertainty (UNC) components for a spatially aggregated quantitative assessment.  We compute the decomposition \eqref{eq:decomposition} at each grid point based on forecasts in the peak monsoon season JAS from all nine evaluation years, and the score components are then averaged across grid points.  The miscalibration--discrimination (MCB--DSC) plots for the mean BS (Figure~\ref{fig:deco}a) and mean CRPS (Figure~\ref{fig:deco}b) provide a spatially consolidated comparison of the forecast methods.  In both cases, the climatological MPC forecast shows the lowest MCB and the lowest DSC component.  The ECMWF raw ensemble (EPS) has higher MCB than all other methods, and the miscalibration is taken care of by post-processing (EPS+ISO, EPS+EMOS).  As regards the statistical forecasts, the inclusion of the ERA5 predictors (Logit-full, DIM-full) models in addition to the correlated rainfall predictors (Logit-basic, DIM-basic) improves DSC while MCB remains similar.  The superiority of the CNN+EasyUQ forecast stems from its elevated discrimination ability.  The Hybrid forecast shows slightly improved skill relative to CNN+EasyUQ, and trades better calibration for even higher discrimination ability.  These findings are stable and apply across all five seasons (Appendix \ref{app:figures}, Figures~\ref{fig:deco_bs} and \ref{fig:deco_crps}).

 \section{Conclusions}  \label{sec:conclusion}

In this work the predictability of one-day ahead, 24-hour precipitation occurrence and amount over northern tropical Africa is investigated on the basis of conventional and new data-driven tools.  Our study builds on previous papers with focus on forecasting rainfall occurrence for the summer season JAS, which compared the performance of climatological, raw and postprocessed ECMWF ensemble forecasts, and a simple logistic regression model based on correlated rainfall predictors.  This binary forecasting problem is revisited in this paper with major adaptions.  Instead of TRMM, GPM IMERG is used as ground truth data source.  Forecasts are produced for the entire year instead of just the summer season (JAS) and ERA5 predictor variables are used to augment the logistic regression model.  To this end, an extensive analysis of weather variables from ERA5 is performed to investigate and understand their relation to and their influence on precipitation.  The meteorological interpretation of these dependencies is obtained by combining previously conducted research and results from statistical analysis performed in this work. 

A key contribution of our work is that we additionally investigate the more challenging problem of producing probabilistic forecasts for accumulated precipitation.  Since the climatology and the NWP model output in this paper are in the form of ensembles, they can be readily used as probabilistic forecasts for precipitation amount.  To produce data-driven statistical forecasts, the Distributional Index Model (DIM) is introduced, which is simple but very effective and thus can serve as a persuasive baseline.  To account for the recent rise of machine learning in weather forecasting, a CNN model is presented which has the additional benefit of inherently exploiting spatial relations.  To obtain a probabilistic output, we couple the CNN model with the recently introduced EasyUQ approach, to yield the CNN+EasyUQ technique.  These different forecasting approaches provide a detailed forecasting benchmark covering the range of simple to sophisticated models and ideas from NWP, statistics, and machine learning in an unprecedented way.  

The CNN+EasyUQ technique outperforms its competitors by a large margin, except for the Hybrid forecast, which is a simple arithmetic average of the HRES+EasyUQ and CNN+EasyUQ forecast distribution that yields minor further improvement.  It is interesting to place our results for one-day ahead, 24-hour forecasts in the context of recent advances in data-based precipitation forecasts.  For nowcasts at prediction horizons up to 12 hours, progress has been persuasive \citep{Ayzel2020, Lagerquist2021, Ravuri2021, Espeholt2022, Zhang2023}.  In stark contrast, recent developments in neural network based weather forecasts at prediction horizons of days ahead have provided scant only attention to rainfall \citep{Bi2023, rasp2023}, arguably due to the recognition that ``precipitation is sparse and non-Gaussian'' \citep[p.~6]{Lam2023}.  The CNN+EasyUQ technique provides an elegant and computationally highly efficient way of addressing the non-Gaussianity of precipitation accumulation.  In very recent work, \citet{andrychowicz2023} find that the data-driven MetNet-3 approach outperforms the ECMWF and NOAA raw ensembles in terms of CRPS for hourly precipitation accumulation over the continental United States at lead times up to 20 hours, but not beyond.  However, unlike our study, which compares the CNN+EasyUQ forecast with state of the art competitors, \citet{andrychowicz2023} do not compare MetNet-3 to postprocessed NWP ensemble forecasts, nor to statistical forecasts of the type considered in our paper. 

The reproduction of the results in this paper requires access to GPM IMERG precipitation data, predictor variables from ERA5, and ECMWF NWP forecasts.  The first two sources are freely accessible, which makes results for MPC, the statistical approaches (Logit and DIM), and our key innovation, the CNN+EasyUQ technique, readily reproducible.  For the more elaborate CNN+EasyUQ approach, code in Python \citep{Python} is available at \url{https://github.com/evwalz/precipitation}.  The raw ECMWF EPS, the postprocessed versions thereof, the HRES+EasyUQ forecast, and the Hybrid model require access to ECMWF NWP forecasts which are freely available using the TIGGE (The International Grand Global Ensemble) archive \citep{TIGGE} instead of MARS from ECMWF.   

In view of its outstanding performance in this study, the CNN+EasyUQ approach can likely improve operational probabilistic forecasts of day ahead, 24-hour rainfall in northern tropical Africa.  To make real-time forecasts feasible, one would need to use the IMERG Early Run (\url{https://gpm.nasa.gov/taxonomy/term/1357}) in lieu of IMERG, which is an option that remains to be tested.   To obtain ensemble forecasts of entire, spatio-temporally coherent precipitation fields, rather than forecasts at individual locations and fixed prediction horizons, the HRES+EasyUQ and CNN+EasyUQ approaches can be coupled with empirical copula techniques \citep{Clark2004, Schefzik2013}, for which we encourage follow-up studies. 

While our study is limited in geographic scope, we feel that data-driven approaches of this type have potential to revolutionize rainfall forecasts throughout the tropics.  Furthermore, the results of comparative studies by \citet{Little2009} for the United Kingdom and \citet{andrychowicz2023} for the continental United States admit the speculation that the CNN+EasyUQ technique can improve probabilistic forecasts of 24-hour precipitation in the extratropics as well.  Finally, a very interesting and relevant research question is whether similar advances in predictive performance are feasible at prediction horizons larger than a day ahead. 

\section*{Acknowledgements}

We thank Sebastian Lerch and Marlon Maranan for comments and discussion. The work of Eva-Maria Walz was funded by the German Research Foundation (DFG) through grant number 257899354.  Tilmann Gneiting is grateful for support by the Klaus Tschira Foundation. 

\bibliographystyle{ametsoc2014}
\bibliography{refs}

\clearpage

\renewcommand{\thefigure}{A\arabic{figure}}
\setcounter{figure}{0}

\appendix 

\section{Additional Figures}  \label{app:figures}
 
\begin{figure}[h]
\centering
\includegraphics[width = \textwidth]{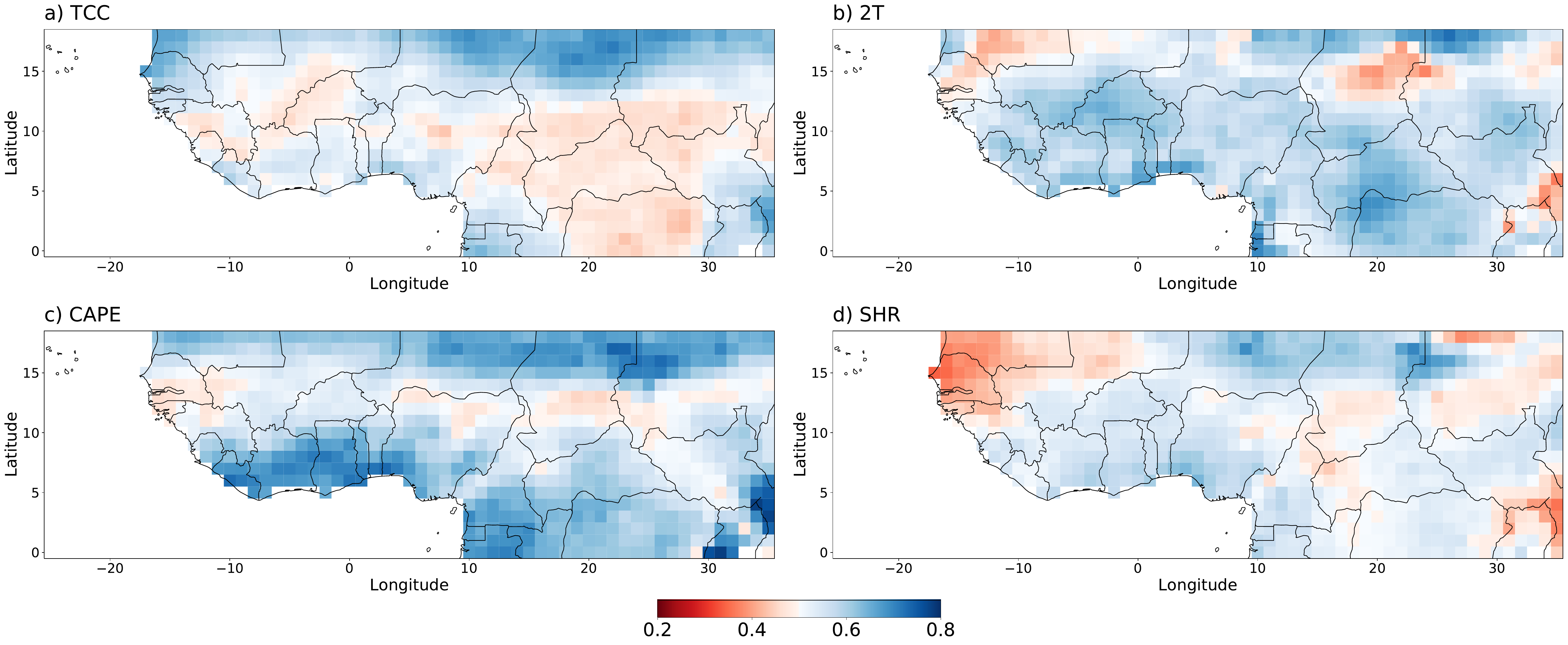}
\caption{As Figure~\ref{fig:CPA_JAS} but for a) TCC, b) 2T, c) CAPE, and d) SHR.  \label{fig:CPA_JAS_2}}
\end{figure}

\begin{figure}[h]
\centering
\includegraphics[width = \textwidth]{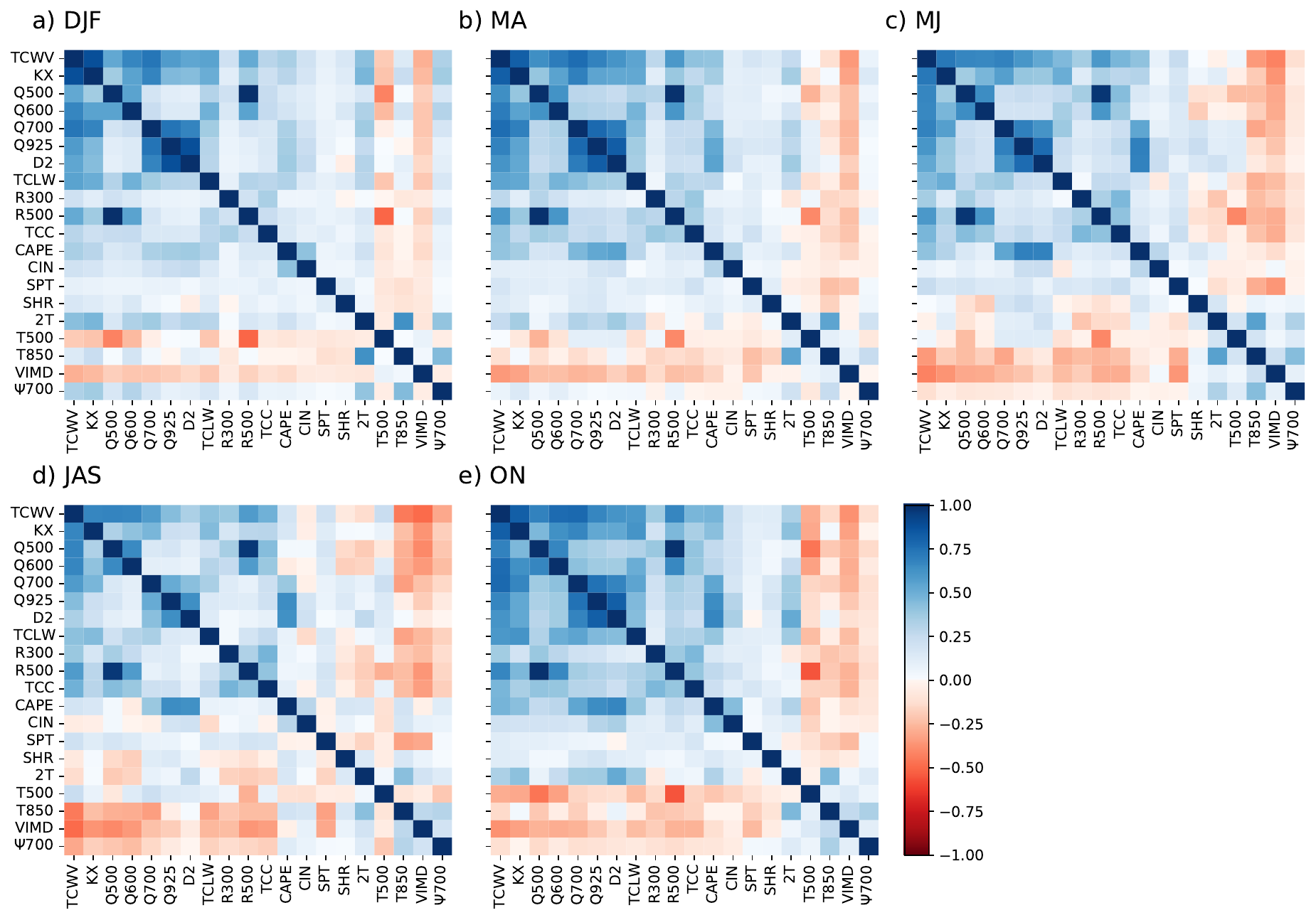}
\caption{As Figure~\ref{fig:spear_mat_jas} but in season a) DJF, b) MA, c) MJ, d) JAS, and e) ON.  \label{fig:spear_mat}} 
\end{figure}

\clearpage

\begin{figure}[p]
\includegraphics[width = \textwidth]{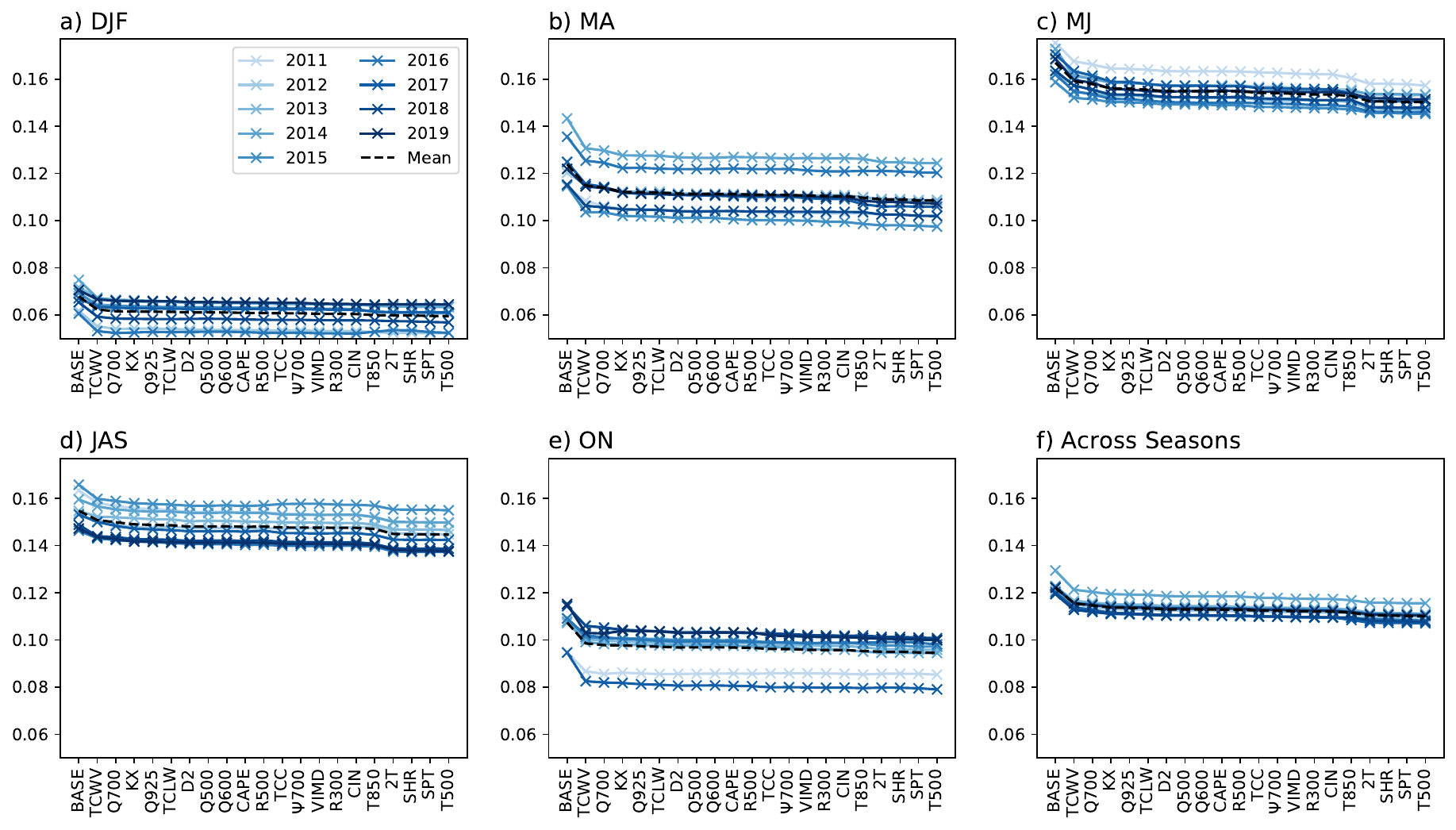}
\caption{As Figure~\ref{fig:feature_sel}a for the Logit PoP forecast, but in season a) DJF, b) MA, c) MJ, d) JAS, and e) ON, and f) across seasons for evaluation folds from 2011 to 2019.  \label{fig:feature_sel_}}
\end{figure}

\begin{figure}[p]
\includegraphics[width = \textwidth]{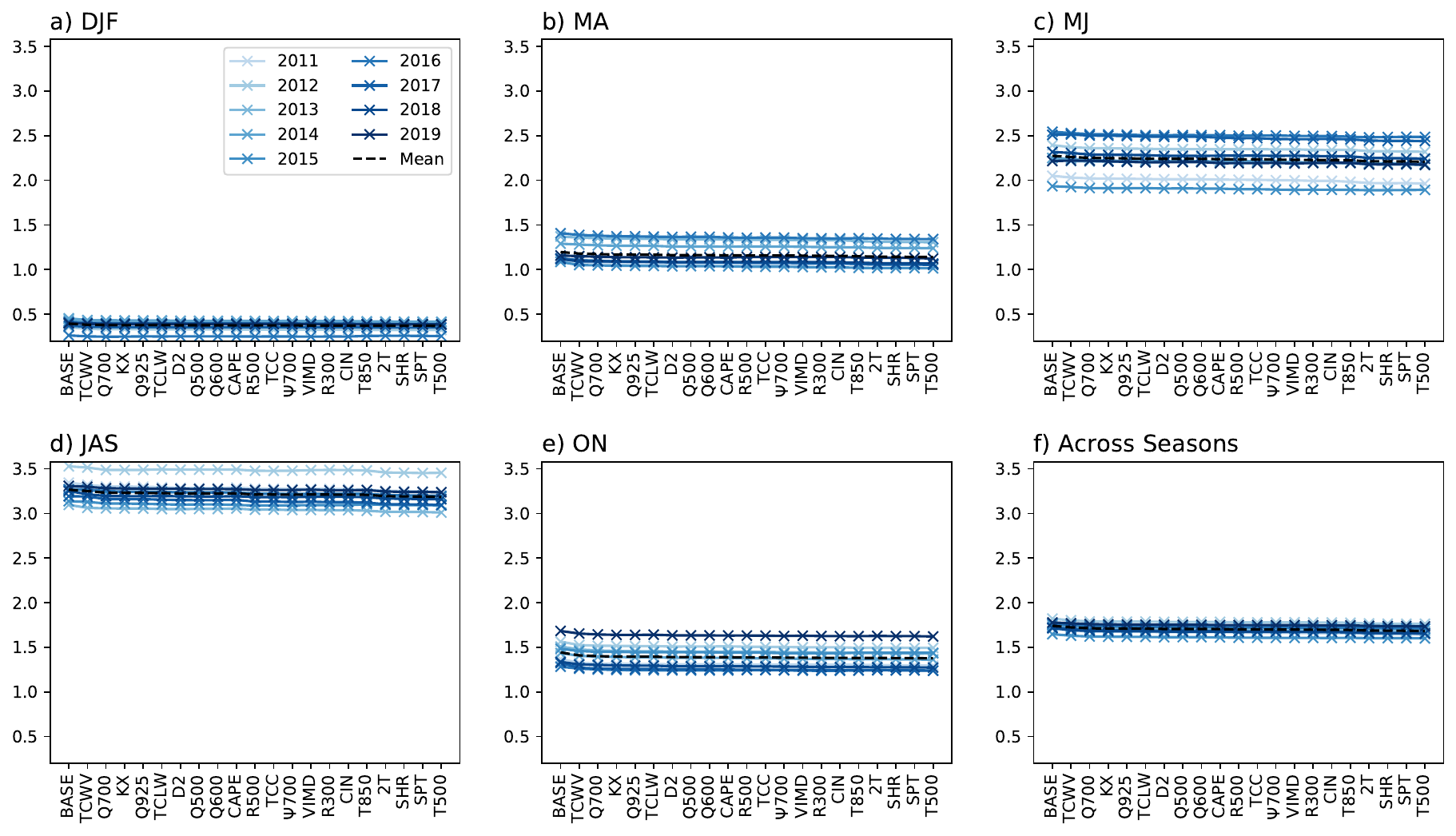}
\caption{As Figure~\ref{fig:feature_sel}b for the DIM forecast of precipitation accumulation, but in season a) DJF, b) MA, c) MJ, d) JAS, and e) ON, and f) across seasons for evaluation folds from 2011 to 2019, in the unit of millimeters.  \label{fig:feature_sel__}}
\end{figure}

\clearpage

\begin{figure}[p]
\centering
\includegraphics[width = \textwidth]{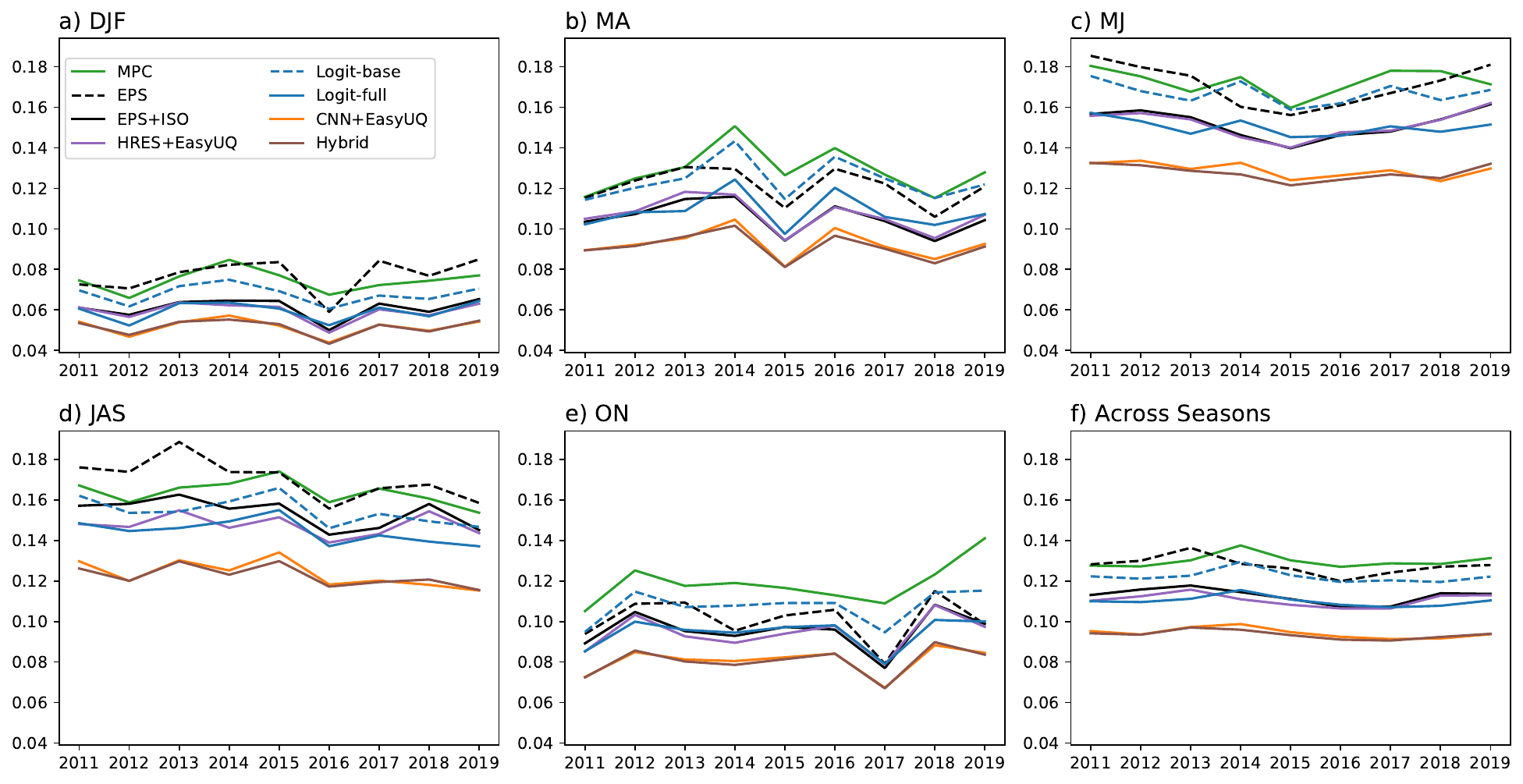}
\caption{Mean Brier score (BS) for PoP forecasts from Table \ref{tab:fct} in season a) DJF, b) MA, c) MJ, d) JAS, e) ON, and f) across seasons, for evaluation folds from 2011 to 2019.  \label{fig:seasonal_bs_mean}}
\end{figure}

\begin{figure}[p]
\centering
\includegraphics[width = \textwidth]{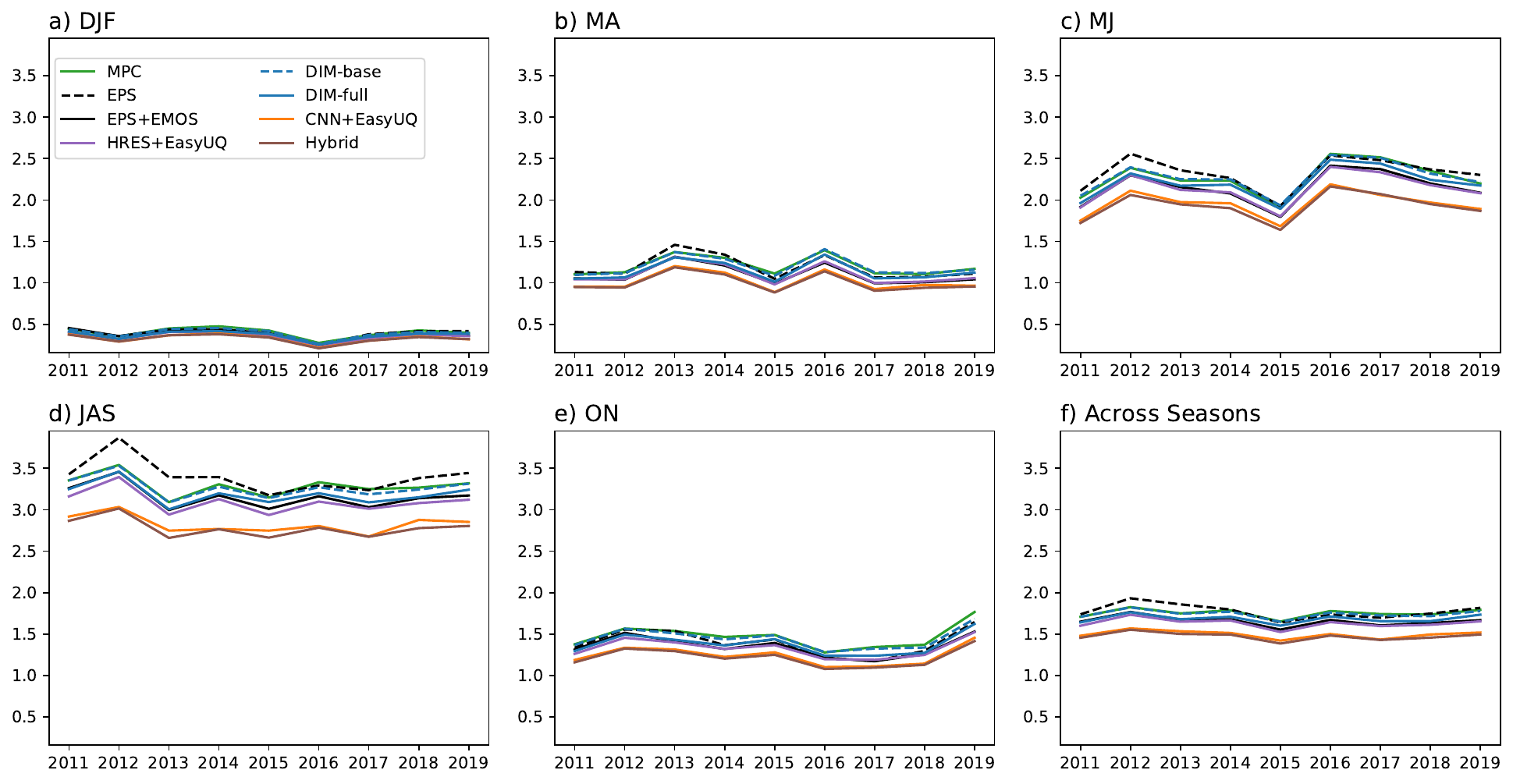}
\caption{As Figure~\ref{fig:seasonal_bs_mean} but for the continuous ranked probability score (CRPS) and probabilistic forecasts of precipitation accumulation in the unit of millimeters.  \label{fig:seasonal_crps_mean}}
\end{figure}

\clearpage

\begin{figure}[p]
\centering
\includegraphics[width = \textwidth]{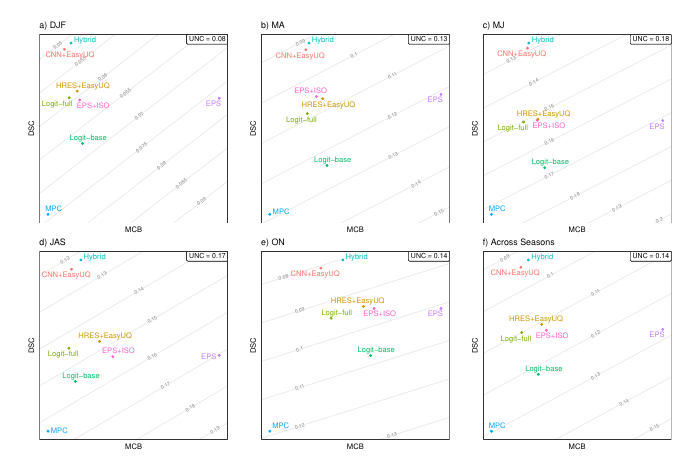}
\caption{Miscalibration (MCB), discrimination (DSC), and uncertainty (UNC) component of the mean Brier score (BS) in season a) DJF, b) MA, c) MJ, d) JAS, e) ON, and f) across seasons for PoP forecasts from Table \ref{tab:fct}.  The CORP decomposition of \citet{Dimitriadis2021} is applied at each grid point, based on the combined evaluation folds from 2011 to 2019, and the mean score and the score components are then averaged over grid points.  Parallel lines correspond to equal mean scores.  \label{fig:deco_bs}}
\end{figure}

\begin{figure}[p]
\centering
\includegraphics[width = \textwidth]{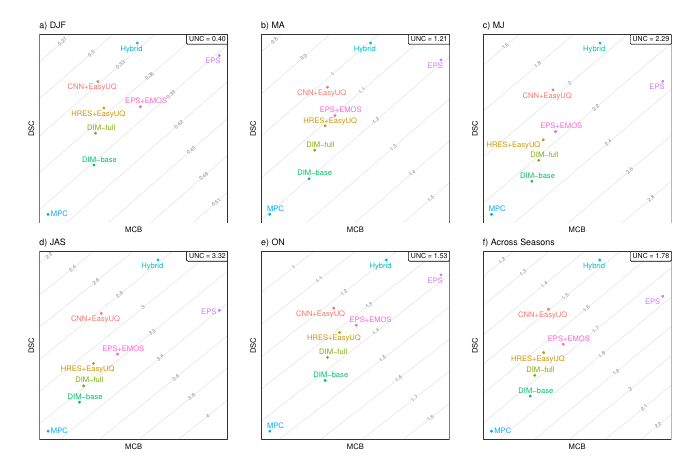}
\caption{As Figure~\ref{fig:deco_bs} but for the mean continuous ranked probability score (CRPS) and probabilistic forecasts of precipitation accumulation in the unit of millimeters.  The isotonicity-based decomposition of \citet{Arnold2023} is used.  \label{fig:deco_crps}}
\end{figure}

\end{document}